\documentclass[aps,showpacs,superscriptaddress,epsf,twocolumn]{revtex4-1}
\usepackage{graphicx}
\usepackage{dcolumn}
\usepackage{natbib}
\usepackage[T1]{fontenc}
\usepackage[utf8]{inputenc}
\usepackage{tensor}
\usepackage{amsmath, amscd, amsthm, amssymb, mathrsfs,amsfonts}
\usepackage{amssymb}
\usepackage{bm}
\newcommand{\mybar}[1]{\setbox0\hbox{{#1}}%
\makebox[\the\wd0][c]{%
\rule[0.42\ht0]{0.75\wd0}{0.7pt}}\hspace*{-\the\wd0}{#1}}
\usepackage{float}
\usepackage{hyperref}
\hypersetup{colorlinks=true, citecolor=blue, urlcolor=blue}
\usepackage{subfigure}
\usepackage{bm}
\usepackage{epstopdf}
\usepackage{amsthm}
\usepackage{float}
\graphicspath{{./figures/}}
\usepackage[toc]{appendix}
\usepackage{color,soul}
\usepackage[dvipsnames]{xcolor}
\usepackage[normalem]{ulem}

\def\A{{\cal A}}

\def\X{{\mathbf X}}

\def\e{{\mathbf e}}

\def\x{{\mathbf x}}
\def\y{{\mathbf y}}

\def\sm1{{\small 1}}
\def\sm2{{\small 2}}

\def\beq{\begin{equation}}
\def\eqn{\end{equation}\noindent}

\begin{document}
\title{{\color{black}Dynamics of coupled $D$-dimensional Stuart-Landau oscillators}}
\author{Pragjyotish Bhuyan Gogoi} 
\affiliation{Department of Physics and Astrophysics, University of Delhi, Delhi 110007, India}
\author{Awadhesh Prasad}
\affiliation{Department of Physics and Astrophysics, University of Delhi, Delhi 110007, India}
\author{Aryan Patel} 
\affiliation{Raphe mPhibr Pvt. Ltd., Noida, Uttar Pradesh 201305, India }
\author{Ram Ramaswamy}
\affiliation{Department of Physical Sciences, IISER, Berhampur, Odisha 760003, India}
\author{Debashis Ghoshal} 
\affiliation{School of Physical Sciences, Jawaharlal Nehru University, New Delhi 110067, India}
\begin{abstract}
The Stuart-Landau oscillator generalized to $D > 2$ dimensions has SO($D$) rotational symmetry. 
We study the collective dynamics of a system of $K$ such oscillators of dimensions $D =$ 3 and 4, 
with coupling chosen to either preserve or break rotational symmetry. This leads to emergent 
dynamical phenomena that do not have analogs in the well-studied case of $D=2$. Further, the 
larger number of internal parameters allows for the exploration of different forms of heterogeneity
among the individual oscillators. 
When rotational symmetry is preserved there can be various forms of synchronization as well as 
multistability and {\em partial} amplitude death, namely, the quenching of oscillations within 
a subset of variables that asymptote to the same constant value. The oscillatory dynamics in 
these cases are characterized by phase-locking and phase-drift. When the coupling breaks 
rotational symmetry we observe {\em partial} synchronization (when a subset of the variables 
coincide and oscillate) and {\em partial} oscillation death (when a subset of variables asymptote 
to different stationary values), as well as the coexistence of these different partial quenching phenomena.  
    
\end{abstract}

\maketitle

\section{\label{sec:1}Introduction}{\color{black}
In the vicinity of a Hopf bifurcation \cite{Pant1,Pikovsky}, the normal form for a generic oscillator 
is the two-dimensional Stuart-Landau (SL) system, 
\begin{equation}
\dot{z} = (\varrho +i\omega -|z|^2)z.
\label{eq:1}
\end{equation}
The equation of evolution above is closely related to the complex Ginzburg-Landau equation, with
$z$ being a complex variable, and $\varrho$ and $\omega$ being real.} When $\varrho \le 0$, 
the origin $z=0$ is an attracting fixed point of the system, while for $\varrho > 0$ the origin
is unstable and the dynamics is instead attracted to a limit cycle, a circle of radius $|z|=\sqrt{\varrho}$. 
There is a supercritical Hopf bifurcation at $\varrho=0$. The equation transforms covariantly under rotation: 
$z\to ze^{i\alpha}$, for a constant angle $\alpha$, 
and since it is exactly solvable, its  simplicity has made it indispensable in the study of 
collective phenomena across diverse fields \cite{Pikovsky, Sune, GarciaMK, Pant1, Boccaletti}. 

Higher dimensional models may be required to describe real-world phenomena such as swarming 
dynamics seen in living organisms and therefore, it is of interest to ask how this model (and 
indeed other such models) can be extended to higher dimensions. To some extent this is a matter 
of choice, and different generalizations of oscillator models, including the Kuramoto phase 
oscillator and its variants to higher dimensions have been proposed recently and their 
collective dynamics have also been studied \cite{Saber,Zhu2,Ott,Boccaletti2,Kumar}.

One of the striking properties of the SL system is its rotational invariance, and keeping this in
mind we have recently proposed a framework that pays special attention to the symmetries of 
Eq.~(\ref{eq:1}) while generalizing it to higher dimensions  \cite{GGGPR}. This is motivated and executed through the use of 
Clifford's geometric algebra \cite{HestenesSob,Doran}, and results in the $D$-dimensional 
generalized SL oscillator that has the following equation of motion:
\begin{equation}
\dot{\mathbf{x}} = (\varrho - r^2) \mathbf{x} + \mathbf{x}\cdot\boldsymbol{\mu},
\label{eq:DdimSL}
\end{equation}
where $\mathbf{x}\in\mathbb{R}^D$ is the coordinate of the oscillator and 
$r = |\mathbf{x}|$ its magnitude. 
In terms of a set of orthonormal basis vectors $\{\mathbf{e}_j\}$, $j=1,2,...,D$ of $\mathbb{R}^D$, 
$\mathbf{x} = \sum x_j \mathbf{e}_j$, the parameters $\mu_{jk}$ define a bivector $\boldsymbol{\mu} = 
\sum_{j < k} \mu_{jk} \mathbf{e}_k \wedge \mathbf{e}_j$, where $\mathbf{e}_k \wedge \mathbf{e}_j = \tfrac{1}{2}
(\mathbf{e}_k\mathbf{e}_j - \mathbf{e}_j\mathbf{e}_k)$ is the antisymmetric wedge product of two vectors in 
Clifford algebra. Since $\mathbf{x}\cdot\boldsymbol{\mu} = \sum_{j<k} \mu_{jk} (x_k \mathbf{e}_j - x_j\e_k)$ 
defines a vector, the above equation transforms covariantly under rotation in $D$-dimensions \cite{GGGPR}. The (real) parameters $\mu_{jk}$ ($j<k$) and $\varrho$ of the generalized 
oscillator correspond to the parameters of rotations (angles) and the radius of the limiting sphere in $D$-dimensions, respectively. The generalized SL oscillator, Eq.~\eqref{eq:DdimSL}, can be solved exactly in all 
dimensions. 

Here we study the dynamics of $K$ coupled oscillators in $D$ dimensions, mainly
$D=3$ and $D=4$ since these are prototypes of odd and even dimensions respectively. (A 
reduction of the $D=4$ case yields quaternionic oscillators that describe three-dimensional 
dynamics.) We will focus primarily on the case of $K=2$ 
oscillators for convenience and simplicity since even this already involves many parameters. 
Our focus is on symmetry, and thus we 
study the dynamics resulting from representative coupling forms that preserve rotational symmetry,
or break the SO($D$) symmetry either partially or completely.

{\color{black} The dynamics of coupled nonlinear oscillators in the vicinity of different bifurcations has been extensively 
studied and analysed \cite{Holmes}. The effects of different forms of couplings \cite{Kurths,Hens,Rajat,Prag,Konishi}
has been one focus of interest since this can affect the dynamics near bifurcations. The behavior of  generic coupled
oscillators near Hopf bifurcation has been modeled by coupled two-dimensional SL oscillators \cite{Aronson}.} Indeed,
the study of ensembles of SL oscillators has been essential in exploration of collective phenomena such as oscillation 
quenching \cite{Kurths, Prag} 
and aging \cite{Vaibhav}, as well as various forms of synchronization \cite{Aronson}, including recent 
discoveries such as explosive synchronization \cite{Gomez,Jafri} and explosive death \cite{Bi}. 

Beyond the widely observed phenomenon of synchronization \cite{Pikovsky,Boccaletti}, its variants and extensions 
\cite{Rosenblum,Pazo,Boccaletti} that can be seen in diverse physical situations 
\cite{Pisarchik,Colet1,Colet2,Makarenko,Tass2}, other important collective phenomena include oscillation 
quenching \cite{Gallego} (which is useful in engineering applications \cite{Kunts,Bode,Marek}) as well as 
amplitude death (AD) \cite{Saxe1,Koseka} and oscillation death (OD) \cite{Koseka,Prag}. The emergence of 
multiple domains of collective behavior in coupled systems is enabled by heterogeneity between 
the individual units  \cite{Kurths,Aronson}. In higher dimensions, these models naturally have a larger number of 
independent internal parameters, and thus a wider variety of heterogeneity can be explored; this leads to 
qualitatively richer collective dynamics. Earlier studies of two-dimensional oscillators show that
breaking the rotational symmetry through the couplings \cite{Nirmal,Wang} can also alter the 
dynamics significantly. The $\tfrac{1}{2} D(D-1)$ dimensional parameter space of the rotation group 
SO($D$) in $D$ dimensions offers the possibility of breaking the rotational symmetry in many different ways, 
providing an insight into the role of symmetry breaking. 

We show here that the interplay of heterogeneity and (broken) symmetry in the couplings leads 
to interesting behavior. In addition to {\em partial synchronization} and {\em partial amplitude death}, where 
only a subset of variables undergo synchronization or amplitude death, respectively, we observe {\em partial oscillation death} where the subset of variables involved asymptote to {\em different} constant values. 
Such partial synchrony or partial quenching has been seen earlier only when there is 
time-delayed interactions \cite{Umeshkanta} or in conjugate-coupled chaotic oscillators \cite{Liu,Chen}.

In Sec. II we briefly recapitulate the analysis of a single oscillator in odd and even dimensions 
and introduce the general coupled oscillator model that will be the focus of our study. In Sec. III we 
investigate the dynamics of two three- and four-dimensional generalized Stuart-Landau oscillators with 
symmetry-preserving coupling, examining phenomena such as multistability and bifurcation. In Sec. IV we extend 
the analysis to symmetry-breaking coupling, and finally summarize our results in Sec. V.

\section{$D$-dimensional Stuart-Landau oscillators}\label{sec:SLD}
The $D$-dimensional Stuart-Landau oscillator~\cite{GGGPR} is covariant under $D$-dimensional rotation
as can be seen by examining Eq.~(\ref{eq:DdimSL}). It is possible to choose a rotated coordinate system 
in which the equations of motion are simpler. 

The bivector $\boldsymbol{\mu}$ has a representation in 
the form of a skew-symmetric matrix $\mathsf{M}$ and can be brought into its normal form through a 
rotation $\x\rightarrow \x\, \mathsf{R} = \y$, where $\mathsf{R}\in\text{SO}(D)$ is a $D\times D$ matrix.
The normal form $\widetilde{\mathsf{M}}$ consists of $N = \lfloor{D/2}\rfloor$ skew-symmetric 2$\times$2  blocks 
$\left(\begin{array}{cc} 0 & -\omega_j\\
\omega_j & 0\end{array}\right)$, $j=1,..., N$ 
for even $D$ dimensions; for odd $D$ there is an additional row and column of zeros. {\color{black}For example, in $D=3$,
$\widetilde{\mathsf{M}}=
\begin{pmatrix}
0 & -\omega_1 & 0\\
\omega_1 & 0 & 0\\
0 & 0 & 0
\end{pmatrix}$ and in $D=4$, $\widetilde{\mathsf{M}}=
\begin{pmatrix}
0 & -\omega_1 & 0 & 0\\
\omega_1 & 0 & 0 & 0\\
0 & 0 & 0 & -\omega_2\\
0 & 0 & \omega_2 & 0
\end{pmatrix}$.} This effectively 
decouples the dynamics of the pairs $y_{2j-1}+iy_{2j} = z_j$ for different $j$. 

Each pair (plus the additional unpaired coordinate 
for odd $D$) evolve independently, subject to the constraint $r=|\mathbf{x}|$. The trajectory is
asymptotically restricted to be on an $N$-dimensional torus $\mathbb{T}^{N} = \mathbb{S}^1_{(a_1)} 
\times \cdots \times \mathbb{S}^1_{(a_N)}$, 
where $a_{(j)}$, determined by initial conditions, is the radius of the circle $\mathbb{S}^1_{(a_j)}$ on the 
$z_j$-plane. (The torus itself is a subset of the hypersphere $\mathbb{S}^{D-1}$ defined by the asymptotic 
value $r^2 \stackrel{t\to\infty}{\longrightarrow} \varrho$.) The ratios of the $\omega_j$'s being 
irrational (or rational, respectively) determine whether the trajectory is space-filling on the torus (or not).  

The asymptotic trajectory in $D=3$ is a circle $\mathbb{T}^{1}\equiv\mathbb{S}^1$. An alternative description is 
possible in this case since the bivector $\boldsymbol{\mu}$ is dual to the vector 
$\boldsymbol{\Omega} = -\e_1 \mu_{23} - \e_2 \mu_{31} - \e_3 \mu_{12}$, which lets us rewrite Eq.~\eqref{eq:DdimSL} as
\begin{equation} 
\Dot{\x}=(\varrho-|\x|^2)\x +\boldsymbol{\Omega} \times \x.
\label{eq:12}
\end{equation}

The second term on the RHS is the usual cross-product of vectors in three dimension. The asymptotic motion is a 
(counter-)clockwise orbit on a limiting circle $\mathbb{S}^1$, with angular frequency 
$|\boldsymbol{\Omega}| = \Omega = \sqrt{ \mu_{12}^2 + \mu_{23}^2 + \mu_{13}^2}$. The circle is centered around 
the vector $\boldsymbol{\Omega}$ on the surface of the invariant sphere $\mathbb{S}^2$ defined by $|\x|^2 = \varrho$ \cite{Ott, GGGPR}.

In $D=4$, for a generic choice of the six independent parameters $\mu_{jk}$ the ratio of the resulting 
independent frequencies $\omega_1$ and $\omega_2$ will be irrational.
Therefore the trajectory for a typical initial condition will be quasiperiodic on the torus ${\mathbb{T}^2}$. 
Since the rotation group SO(4)$\sim$ SO(3) $\times$ SO(3), a special reduction is possible here: if  
one of the following restrictions is imposed on the components $\mu_{ij}$,
\begin{equation}
\mu_{i4} = \pm \frac{1}{2} \sum_{jk} \epsilon_{ijk} \mu_{jk} \equiv \nu_{i},\: (i,j,k=1,2,3),
\label{eq:Q}
\end{equation}
the constrained bivector $\boldsymbol{\mu} = \nu_1(\e_1\e_4 \pm\e_2\e_3) + \nu_2(\e_2\e_4 \pm 
\e_1\e_3) + \nu_3(\e_3\e_4 \pm \e_1\e_2)$ has only three independent parameters. The combinations accompanying 
$\nu_i$ are precisely the imaginary units of quaternions. It results in a circular periodic (rather than 
quasiperiodic, since the two frequencies become identical \cite{GGGPR}) trajectory that has the symmetries of a 
three-dimensional system.

For an ensemble of $K$ coupled generalized SL oscillators, keeping the interactions pairwise, the equations 
of motion are 
\begin{equation}
\dot{\mathbf{x}}_{n} = (\varrho_{n} - r_{n}^2) \mathbf{x}_{n} + \boldsymbol{\mu}^{(n)}\cdot\mathbf{x}_{n} + 
\sum_{m=1}^{K} \varepsilon_{nm} \A_{nm} \mathbf{g}(\mathbf{x}_{n}, {\mathbf x}_{m}),
\label{eq:EomNCpldOsc1}
\end{equation}

where the subscripts $n, m = 1,2,..., K$ refer to the oscillators. We will use, when required, 
superscripts to indicate the components of the vectors, namely $\x_{n}=(x_n^1, x_n^2,..., x_n^D)$. 
The oscillator indices are denoted by superscripts on the bivectors. The elements ${\A}_{nm}$ of the 
adjacency matrix are equal to 1, if the oscillators $n$ and $m$ are coupled, and zero otherwise. The 
coupling strengths and its functional form are denoted by $\varepsilon_{nm}$ and $\mathbf{g({x}}_n, \mathbf{x}_m)$, respectively. 
By choosing different functions $\mathbf{g}$ the rotational symmetries  of the system may either be 
preserved or broken, partially or entirely. Evidently there are many parameters that can be 
varied to explore a rich landscape of possible dynamics.

\section{SO($D$) preserving coupling \label{sec:3}}
The simple choice of linear diffusive coupling in which $\mathbf{g}(\mathbf{x}_n, \mathbf{x}_m) = 
\mathbf{x}_m - \mathbf{x}_n$, leads to
\begin{equation}
\dot{\mathbf{x}}_n = (\varrho_n - r_n^2) \mathbf{x}_n + \mathbf{x}_n\cdot\boldsymbol{\mu}^{(n)} + 
\sum_{m}^{K} \varepsilon_{n,m} (\mathbf{x}_m - \mathbf{x}_n)
\label{eq:EomNCpldOsc2}
\end{equation}
for $n=1,..., K$. This preserves covariance since the coupled equations transform in the same manner 
under rotations of the coordinate system. Such coupling has also been termed {\em scalar} \cite{Aronson}. 

Let $\mathsf{R}_n$ be the orthogonal $D\times D$ matrix that brings the matrix $\mathsf{M}^{(n)}$ 
(corresponding to $\boldsymbol{\mu}^{(n)}$) to its normal form $\mathsf{R}_n \mathsf{M}^{(n)} \mathsf{R}_n^T = \widetilde{\mathsf{M}}^{(n)}$. 
The equation of motion of the $n$th oscillator, when multiplied from the {\em right} by $\mathsf{R}_n$ 
to define $\mathbf{y}_n=\mathbf{x}_n\mathsf{R}_n$, takes the form
\begin{equation}
\dot{\mathbf{y}}_n = (\varrho_n-r_n^2) \mathbf{y}_n + \mathbf{y}_n\cdot\widetilde{\boldsymbol{\mu}}^{(n)}
+ \sum_{m\ne n} \varepsilon_{n,m} (\mathbf{y}_n -   \mathbf{y}_m \mathsf{R}_m^T \mathsf{R}_n).
\label{eq:EomNCpldOscNormal}
\end{equation}

One should hasten to add that while for any of the individual oscillators, the rotation $\mathsf{R}_n$ 
would convert the coordinate $\mathbf{x}_n$ to the corresponding normal coordinate $\mathbf{y}_n$, 
the latter in the above equations are merely linear combinations of the original degrees of freedom. 
This is because the `rotations' for different values of $n$ do not necessarily commute, 
therefore, cannot be made together. Nevertheless, the above form of the equations may be used to 
simplify numerical investigation. 

A notable exception where the transformation is indeed a (single) rotation is for the special case 
where the bivector parameters for the oscillators are proportional to each other, namely
\begin{equation}
\boldsymbol{\mu}^{(n)} = \gamma_n\boldsymbol{\mu},
\label{eq:muprop}
\end{equation} 
where $\gamma_n$ are real constants. In this case the orthogonal matrices $\mathsf{R}_n$ that bring 
the matrices $\mathsf{M}^{(n)}$ to the canonical form are identical, i.e., $\mathsf{R}_n = \mathsf{R}$ 
for all $n$, consequently $\mathsf{R}_n \mathsf{R}_m^T = \mathbf{1}$. Hence Eqs.~\eqref{eq:EomNCpldOscNormal} 
takes the form of Eq.~\eqref{eq:EomNCpldOsc2}, but for the coupled normal coordinates. 
As in the case of a single oscillator these decouple into $N= \lfloor D/2\rfloor$ blocks of size $2\times 2$, and an
additional equation for odd $D$. The equations for the pairs is expressed compactly in
terms of $z_n^j = y_n^{2j-1} + i y_n^{2j}$: 
\begin{align}
\dot{z}_n^j &= (\varrho_n - r_n^2) z_n^j + i \gamma_n \omega_j z_n^j + \sum_{m\ne n} 
\varepsilon_{n,m} (z_n^j - z_m^j) \nonumber\\
{} &= \mathfrak{c}_n^j z_n^j - \sum_{m\ne n} \varepsilon_{n,m} z_m^j,
\label{eq:eomCpldCmplx}
\end{align}
where $\mathfrak{c}_n^j = (\varrho_n + \sum_{m\ne n} \varepsilon_{n,m} - r_n^2) + i \gamma_n \omega_j$.
An equivalent representation is the matrix form $\dot{Z}_j = L_K Z_j$, where the $K\times K$ matrix 
$L_K$ has $\mathfrak{c}_n^j$'s on the diagonal and the off-diagonal $(n,m)$-th 
term is $-\varepsilon_{n,m}$. Note that $L_K$ is not the Jacobian matrix $J_K \sim \partial\dot{z}_i/\partial z_j$ 
since $r_n^2 = \sum_j |z_n^j|^2$ contributes to the latter.

For odd $D$, there is an additional (unpaired) equation
\begin{equation}
\dot{y}_n^D = (\varrho_n - r_n^2) y_n^D + \sum_{m\ne n} \varepsilon_{n,m} (y_n^D - y_m^D).
\label{eq:eomCpldRLast}
\end{equation}
As can easily be deduced from the equations above, if the initial condition is such that $y_n^{D}(t=0) = 0$, 
it remains zero at all times. As a result the dynamics is restricted to the hyperplane 
perpendicular to $\e_D$, reducing it to the $(D-1)$-dimensional case. This is true of both coupled 
as well as uncoupled oscillators.

Although the structure of the equations and the decomposition into the block-diagonal form is very 
similar to the case of a single (uncoupled) oscillator \cite{GGGPR}, the nature of the dynamics will be 
different because there is coupling between the oscillators in each sector. For example, the 
coupling can change the ratios of the frequencies for the different pairs, from rational to 
irrational or vice versa. Thus space filling orbits can, under the influence of coupling, 
become periodic or vice versa. 

In polar coordinates $z_n^j = r_{n,j} e^{i\phi_{n,j}}$, Eqs.~\eqref{eq:eomCpldCmplx} take the form 
\begin{align}
\dot{r}_{n,j} &= \left(\varrho_n + \sum_{m\ne n} \varepsilon_{n,m} - r_n^2\right)\, r_{n,j} \nonumber \\
               & \hspace{2cm}- \sum_{m\ne n} \varepsilon_{n,m} r_{m,j} \cos(\phi_{n,j} - \phi_{m,j}), \nonumber\\ 
\dot{\phi}_{n,j} &= \gamma_n\omega_j + \frac{1}{r_{n,j}} \sum_{m\ne n} \varepsilon_{n,m} r_{m,j} \sin(\phi_{n,j} - \phi_{m,j}). \label{eq:eomCpldRPhi}
\end{align}

If all the frequencies are identical, namely when $\gamma_n=1$, $\phi_{n,j} = \phi_{m,j}$ for all 
$n,m$ is a solution since the terms involving the (co)sine functions in the RHS above vanish, 
decoupling the equations of motion for $r_{n,j}$ and $\phi_{n,j}$. The second set of equations are 
easily integrated to get $\phi_{n,j} = \omega_j t$ for all $n$: the oscillators are in complete 
phase synchronization in this case. 

{\color{black}We now look at some specific cases of $K=2$ oscillators in $D = 3$ and 4, first 
considering couplings that keep the equations covariant.} With some obvious simplification of notation, the equations for the two oscillators in such a case are

\begin{eqnarray}
\dot{\mathbf{x}}_1 &=& (\varrho_1 -r_1^2) \mathbf{x}_1 + \mathbf{x}_1\cdot\boldsymbol{\mu}^{(1)} + \varepsilon_1 (\mathbf{x}_2 - \mathbf{x}_1), \nonumber\\
\dot{\mathbf{x}}_2 &=& (\varrho_2-r_2^2) \mathbf{x}_2 + \mathbf{x}_2\cdot\boldsymbol{\mu}^{(2)} + 
\varepsilon_2 (\mathbf{x}_1 - \mathbf{x}_2),
\label{eq:CpldOsc}
\end{eqnarray}

from which one can obtain the equations of motion for the magnitudes $r_1 = |\x_1|$, 
$r_2 = |\x_2|$ and the relative angle $\alpha = \cos^{-1} \big(\tfrac{\x_1\cdot\x_2}{|\x_1|\,|\x_2|}\big)$:
\begin{align}
\dot{r}_1 &= (\varrho_1 - r_1^2) r_1 - \varepsilon _1 (r_1 - r_2 \cos\alpha),\nonumber\\
\dot{r}_2 &= (\varrho_2 - r_2^2) r_2 - \varepsilon_2  (r_2 - r_1 \cos\alpha),\nonumber  \label{eq:MagAngl}\\
\frac{d}{dt}\cos\alpha &= \sum_{ij} \big( \mu^{(2)}_{ij} - \mu^{(1)}_{ij}\big) \frac{x_1^i}{r_1} 
\frac{x_2^j}{r_2} \\
&\hspace{2cm}+\left(\varepsilon_1 \frac{r_2}{r_1}+ \varepsilon_2 \frac{r_1}{r_2} \right) 
\sin^2\alpha\nonumber.
\end{align}
{\color{black} It should be kept in mind that the symmetries that are being preserved in Eq.~(\ref{eq:CpldOsc})
pertain to the normal form described by SL oscillators, and not necessarily to those in terms of the ``original variables'' of coupled generic oscillators.

When the coupled oscillators in Eq.~(\ref{eq:CpldOsc}) are two-dimensional with $\varrho_1=\varrho_2$ and $\varepsilon_1=\varepsilon_2$, simplification of the 
term $\sum_{ij} \big( \mu^{(2)}_{ij} - \mu^{(1)}_{ij}\big) \frac{x_1^i}{r_1}\frac{x_2^j}{r_2}=(\omega^{(2)}-
\omega^{(1)})\sin{\alpha}$, where $\omega^{(1)}$ and $\omega^{(2)}$ are the characteristic frequencies of the two oscillators, allows for a reduction of Eq.~(\ref{eq:MagAngl}) to a form describing the behavior of 
two generic oscillators, coupled linearly and diffusively, that are near Hopf bifurcation \cite{Aronson}.} 
In the special case $\boldsymbol{\mu}^{(1)} = \boldsymbol{\mu}^{(2)}$, the equations for $r_1$, 
$r_2$ and $\cos\alpha$ in Eq.~(\ref{eq:MagAngl}) form a closed system (in any dimension). The angle $\alpha$ then evolves as
\begin{equation}
\dot{\alpha}= - \left( \varepsilon_1 \frac{r_2}{r_1} +\varepsilon_1 \frac{r_2}{r_1} \right) \sin{\alpha} = - C \sin\alpha, \label{eq:EvolvAngl} 
\end{equation}
where $C>0$, implying the existence of a stable fixed point at $\alpha=0$. Details of this fixed point of Eq.~\eqref{eq:MagAngl} with regard to its coordinates, $(r_{1*}, r_{2*}, 1)$, and its stability are discussed in Appendix \ref{Ap:1}. Since $\cos\alpha \stackrel{t\to\infty}{\longrightarrow} 1$, the two vectors $\x_1$ and $\x_2$ align asymptotically, and their lengths $r_{1*}$ and $r_{2*}$  (which are not equal in general) are constants. In terms of the asymptotic coordinates $\boldsymbol{\xi}_{1,2}(t) = \displaystyle{\lim_{t\to \infty}} \x_{1,2}(t)$, which are free of initial transient behavior, $\boldsymbol{\dot{\xi}}_{1,2} = \boldsymbol{\xi}_{1,2} \cdot \boldsymbol{\mu}$, describing a linear system that undergoes pure rotations. Notably, the coupled dynamics here is characterized by extreme multistability \cite{GGGPR, Showalter}; if we choose any initial $\left(\x_1(0), \x_2(0)\right)$ such that $\left(r_1(0), r_2(0), \cos\alpha (0)\right) = (r_{1*}, r_{2*}, 1)$, these 
values remain constant. For the special case $\varrho_1 = \varrho_2 = \varrho$, the existence of 
the fixed point set $(\sqrt{\varrho},\sqrt{\varrho},1)$ 
suggests the two systems are completely synchronized in the asymptotic limit, occupying 
identical limit cycles. Interestingly, if the initial conditions are such that $r_1(0)=r_2(0)$ (a 
particular choice), the "center of mass" $\X=(\x_1+\x_2)/2$ acts as a single uncoupled $D$-dimensional 
oscillator. The attractors in this case can be determined exactly. 

\subsection{Coupled oscillators in $D=3$}
In three dimensions the duality between bivectors and vectors can be used to express the evolution  
Eq.~\eqref{eq:CpldOsc} of the two coupled oscillators as
\begin{eqnarray}
\dot{\x}_1 &=& (\varrho_1 - r_1^2) \x_1 + {\boldsymbol{\Omega}^{(1)}} \times \x_1 +\varepsilon_1 (\x_2 -\x_1),\nonumber\\
\dot{\x}_2 &=& (\varrho_2 - r_2^2) \x_2 + {\boldsymbol{\Omega}^{(2)}} \times \x_2 +\varepsilon_2 (\x_1 -\x_2),\label{eq:Cpld3D}
\end{eqnarray}
where $\boldsymbol{\Omega}^{(1)}$ and $\boldsymbol{\Omega}^{(2)}$ are the vectors 
that are dual to the bivectors $\boldsymbol{\mu}^{(1)}$ and $\boldsymbol{\mu}^{(2)}$ respectively. 
In the following we explore the diverse forms of dynamics that arise for different choices of $\boldsymbol{\Omega}^{(1)}$ and $\boldsymbol{\Omega}^{(2)}$.\\

\subsubsection{Common eigenvectors}
When the bivectors differ by a scalar, i.e., $\boldsymbol{\mu}^{(2)} = \boldsymbol{\mu}^{(1)}/\gamma$,
then $\boldsymbol{\Omega}^{(2)} = \boldsymbol{\Omega}^{(1)}/\gamma$ and the 
corresponding eigenvectors are parallel. Consider the case with $\varrho_1 = \varrho_2$ 
and $\varepsilon_1 = \varepsilon_2$. 
Due to the exchange symmetry of the oscillators, we only need to consider $\gamma\in (0,1]$ in order to 
explore the effect due to the difference in frequency. Without loss of generality (and for 
convenience) we shall treat the coupled systems in their normal-mode coordinates, in which the equations for the components read [in the following $(n,m)=(1,2)$ or (2,1)]
\begin{align}
&\dot{y}^1_n = [\varrho_n - (r_n)^2] y^1_n - \Omega^{(n)} y^2_n + \varepsilon_n (y^1_m - y^1_n),\nonumber \\ 
&\dot{y}^2_n = [\varrho_n - (r_n)^2] y^2_n + \Omega^{(n)} y^1_n + \varepsilon_n (y^2_m - y^2_n),\nonumber \label{eq:3D2a}\\
&\dot{y}^3_n = [\varrho_n - (r_n)^2] y^3_n + \varepsilon_n (y^3_m - y^3_n),
\end{align}
where $\Omega^{(1)} = |\boldsymbol{\Omega}^{(1)}| \equiv \Omega$ and 
$\Omega^{(2)} = |\boldsymbol{\Omega}^{(2)}| = \Omega/\gamma$ are their characteristic frequencies. 

We continue with the choice $\varrho_1 = \varrho_2 = \varrho$ and $\varepsilon_1 = \varepsilon_2 
= \varepsilon$ for the moment.
In the absence of coupling, the two oscillators share a common axis of rotation along $\e_3$. With 
the introduction of coupling, there appears a special kind of multistability. If initial values are 
chosen such that $y^{3}_1(0) = 0$ and $y^{3}_2(0) = 0$, namely the two oscillators start from their
common equatorial (12)-plane, their trajectories remain confined to this plane at all times. 
The dynamics for these initial conditions thus reduces to that of two scalar-coupled SL oscillators 
in $D=2$, a case previously studied in detail by Aronson et al. \cite{Aronson}.

\begin{figure}[htp]
      \centering
     {\includegraphics[width=0.4\textwidth]{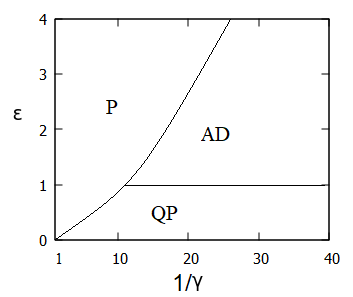}}
     \caption{Schematic phase diagram of two coupled oscillators in $D=2$. 
     For different values of $\varepsilon$ and ${\gamma}$, periodic and quasiperiodic motion, and amplitude death can be observed.
     The boundaries of the different regions are constructed based on the largest two Lyapunov exponents.}
      \label{fig:1}
\end{figure}
 
As shown schematically in Fig.~\ref{fig:1}, the contours of $\lambda_2 = 0$ and 
$\lambda_3 = 0$, corresponding to the vanishing of the two largest Lyapunov exponents, 
divide the $\gamma$-$\varepsilon$ parameter space into three dynamical regions corresponding to 
quasiperiodic (QP) motion (when both exponents are zero), periodic motion (P) (one is zero and the 
other one negative), and amplitude death (AD) (both exponents are negative). As was noted 
\cite{Aronson},  two modes of collective behavior are possible: phase locking when 
the dynamics is periodic, and phase drift when it is quasiperiodic. 

\begin{center}
\begin{figure}[htp]
         \includegraphics[width=0.38\textwidth]{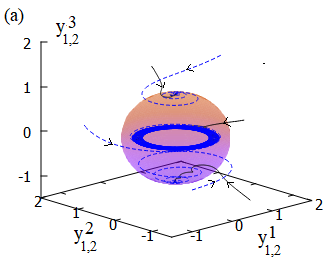}\\ 
     \includegraphics[width=0.38\textwidth]{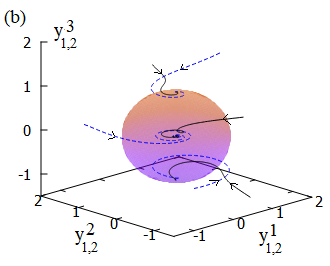}\\
      \includegraphics[width=0.38\textwidth]{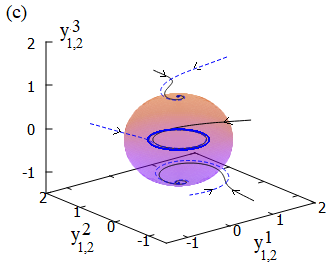}
     \caption{(Color online) Trajectories (solid black and dashed blue lines) of the two coupled three-dimensional oscillators outlining the different forms of multistability present in Eq.~(\ref{eq:3D2a}). For $\gamma = 0.05$, there are three attractors: (a) for $\varepsilon=0.5$ one is quasiperiodic and two are fixed points, (b) for $\varepsilon=1.5$, all three are fixed points, and (c) for $\varepsilon=3$ there are two fixed points and one periodic orbit. (The plots correspond to $\Omega=0.2$ and $\varrho=1$.)}
      \label{fig:2}
\end{figure}    
\end{center}

The equatorial plane in $D=3$ is an invariant sub-manifold that is unstable to transverse 
perturbations for any $\varepsilon>0$. The two stable fixed points at the poles of the 
sphere $(r_{1,2})^2=\varrho$  attract initial values not on the equatorial plane. Thus there are
three kinds of multistability, the $\gamma$-$\varepsilon$ 
domains corresponding to each one of which can be analytically obtained using results 
from~\cite{Aronson}. If we define $\Delta = \Omega(1-\frac{1}{\gamma})$ and consider $\varrho = 1$, 
these are: (1) two AD and one quasiperiodic state [see Fig.~\ref{fig:2}(a)], which occur when 
$|\Delta| < 2$ and $\varepsilon \in (0,\tfrac{1}{2} |\Delta|)$ or when $|\Delta| > 2$ and 
$\varepsilon \in (0,1)$; (2) three AD states [see Fig.~\ref{fig:2}(b)] for $|\Delta|> 2$ and 
$\varepsilon \in (1,\tfrac{1}{2} (1 + \tfrac{1}{4}\Delta^2))$; and (3) two AD and one periodic 
state [see Fig.~\ref{fig:2}(c)] for $|\Delta|<2$ and $\varepsilon > \tfrac{1}{2} |\Delta|$ or for 
$|\Delta| > 2$ and $\varepsilon > \tfrac{1}{2} (1 +\tfrac{1}{4}\Delta^2)$. It can further be shown 
that when the dynamics is quasiperiodic, as in Fig.~\ref{fig:2}(a), the radii enclosing the 
annular projections of the flow over the equatorial plane are $r_\pm^2 = a(1\pm c)$, where $a = 1 - \varepsilon$ 
and $c = {\varepsilon}/{\sqrt{a^2 + \tfrac{1}{4}\Delta^2}}$ (see Appendix \ref{Ap:3} for details). 
Similarly, when the dynamics is periodic, the (equal) radius of the circular limit cycles, one of 
which is shown in Fig.~\ref{fig:2}(c) for the second oscillator, is 
$r_*^2 = 1 - \varepsilon + \sqrt{\varepsilon^2 - \tfrac{1}{4}\Delta^2}$. Additionally, the
oscillators are phase locked at a relative angle $\alpha_* = \sin^{-1}(\Delta/2\varepsilon)$ (see Appendix \ref{Ap:3} for details). For numerical integration of Eq.~(\ref{eq:3D2a}) as well as other equations in the text, we use the RK4 method with a stepsize of 0.01.    

\begin{center}
\begin{figure}[htp]
     {\includegraphics[width=0.35\textwidth]{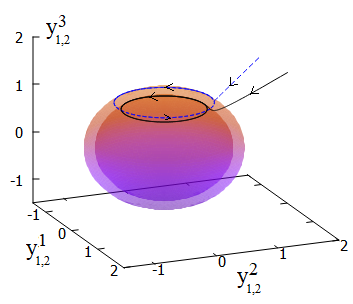}}
     \caption{(Color online) Trajectories (solid black and dashed blue lines) of the two coupled three-dimensional oscillators in Eq.~(\ref{eq:3D2a}) with identical frequency $\boldsymbol{\mu}^{(1)} = \boldsymbol{\mu}^{(2)}$ and parameters 
     $\varrho_1=1$, $\varrho_2=1.5$, $\varepsilon_1=0.5$ and $\varepsilon_2=0.2$. 
     They approach two circular limit cycles located on the spheres of radii $r_{1*}\approx 1.04$ and $r_{2*}\approx 1.21$.}
    \label{fig:3}
\end{figure} 
\end{center}

{\color{black}It is possible to find the fixed points of Eqs.~\eqref{eq:CpldOsc} and analyze their stability when $\gamma = 1$ even with $\varepsilon_1 \ne \varepsilon_2$, 
as is discussed in Appendix~\ref{Ap:1}.} The case of $\varrho_1=1$, $\varrho_2=1.5$
$\varepsilon_1 = 0.5$, and $\varepsilon_2 = 0.2$  for which the fixed points are $r_{1*}\approx 1.04$, $r_{2*} 
\approx 1.21$ and $\cos{\alpha_*} = 1$ is shown in Fig.~\ref{fig:3}. This is a special case of {\em phase synchronized} orbits,  with the position vectors of the two oscillators being 
completely aligned although their lengths, which are fixed asymptotically, are different. 
Different initial conditions lead to different asymptotic trajectories, so this is also a case of 
extreme multistability. {\color{black}Moreover, as also alluded to in Sec.~\ref{sec:3}, there are a few special solutions possible in this case. First, a zero initial relative phase always remains zero, i.e. $\alpha(t)=0$ at all times if $\alpha(0)=0$; the oscillators are phase synchronized at all times in this case. When $\alpha(0)\neq0$, on the other hand, phase synchrony is reached in the asymptotic limit, i.e. $\displaystyle{\lim_{t\to\infty}}\alpha(t)=0$. Similarly, for two coupled odd-dimensional oscillators, initial conditions $y_1^D(0)=y_2^D(0)=0$ lead to $y_1^D(t)=y_2^D(t)=0$ at all times. These, however, are not stable solutions since with slightly different initial conditions, e.g. such that $(y_1^D(0)\neq0,y_2^D{(0)}=0)$, the coupling results in $\displaystyle{\lim_{t\to\infty}}y_1^D(t)=\displaystyle{\lim_{t\to\infty}}y_2^D(t)=\text{constant}\neq0$.} It can further be shown (see Appendix \ref{Ap:4}) that when $\varepsilon_1 = 
\varepsilon_2 = \varepsilon \to \infty$, $r_{1*}, r_{2*} \stackrel{\varepsilon\to\infty}{\longrightarrow} 
\tfrac{1}{2} (\varrho_1 + \varrho_2)$ which implies that the system transitions from a state of phase 
synchronization to complete synchronization with increasing coupling strength.

\medskip

\subsubsection{ Different eigenvectors}
Eigenvectors for generic choices of $\boldsymbol{\Omega}^{(1)}$ and $\boldsymbol{\Omega}^{(2)}$, 
will not be parallel but the normalised eigenvectors can be chosen such that $\e_2$ and $\e_3$ 
lie on the plane defined by $\boldsymbol{\Omega}^{(1)}$ and $\boldsymbol{\Omega}^{(2)}$, with one 
of them, say, $\e_3$, along the bisector of the angle between them. In this basis 
$\boldsymbol{\Omega}^{(1)} = \mu^{(1)}_{13}\e_2 - \mu^{(1)}_{12}\e_3$ and $\boldsymbol{\Omega}^{(2)} = - \frac{1}{\gamma} (\mu^{(1)}_{13}\e_2 + \mu^{(1)}_{12}\e_3)$, with 
$\gamma = {|\boldsymbol{\Omega}^{(1)}|}/{|\boldsymbol{\Omega}^{(2)}|} \equiv |\Omega^{(1)}/\Omega^{(2)}|$. 

For the case $\varrho_1 = \varrho_2 = \varrho$ and $\varepsilon_1 = \varepsilon_2 = \varepsilon$  
Eq.~\eqref{eq:CpldOsc} becomes 

\begin{align}  
&\dot{x}^1_n = [\varrho-(r_n)^2] x^1_n + \mu^{(n)}_{12} x^2_n + \mu^{n}_{13} x^3_n + \varepsilon (x^1_m - x^1_n), \nonumber\\
&\dot{x}^2_n = [\varrho-(r_n)^2] x^2_n - \mu^{(n)}_{12} x^1_n + \varepsilon (x^2_m - x^2_n), \label{eq:3DC3} \nonumber \\
&\dot{x}^3_n = [\varrho-(r_n)^2] x^3_n - \mu^{(n)}_{13} x^1_n + \varepsilon (x^3_m - x^3_n), 
\end{align}

where $(n,m) = (1,2)$ or (2,1) as before, and $\mu^{(2)}_{12}=\mu^{(1)}_{12}/\gamma$ and 
$\mu^{(2)}_{13}=-\mu^{(1)}_{13}/\gamma$. For $\gamma=1$, the above equations are invariant 
under $(x^1_1, x^2_1,  x^3_1) \leftarrow \pm (x^1_2, x^2_2, -x^3_2)$, 
whence the dynamics is confined to the stable manifold defined by the simultaneous equations 
$x^1_1 = x^1_2$, $x^2_1 = x^2_2$ and $x^3_1 = - x^3_2$. It can easily be shown that the coordinates 
$(x_{1*}^1,x_{1*}^2,x_{1*}^3)$ of the fixed points on this manifold are solutions of the quadratic equations

\begin{eqnarray}  
\mu^{(1)}_{12} \big[ (x^1_{1*})^2+ (x^2_{1*})^2 \big] + \mu^{(1)}_{13} x^2_{1*} x^3_{1*} &=& 0, \nonumber\\
x^2_{1*} \big[\varrho - (r_{1*})^2\big] - \mu^{(1)}_{12} x^1_{1*} &=& 0, \label{eq:Ch3DC4} \\
\mu^{(1)}_{12} x^1_{1*} x^3_{1*} - \mu^{(1)}_{13} x^1_{1*} x^2_{1*} - 2\varepsilon x^2_{1*} x^3_{1*} &=& 0. \nonumber
\end{eqnarray}

The fixed points come in pairs $\pm (x^1_{1*}, x^2_{1*}, x^3_{1*})$ and can be obtained 
numerically for various choices of $\varepsilon$. 

\begin{center}
\begin{figure}[htp]
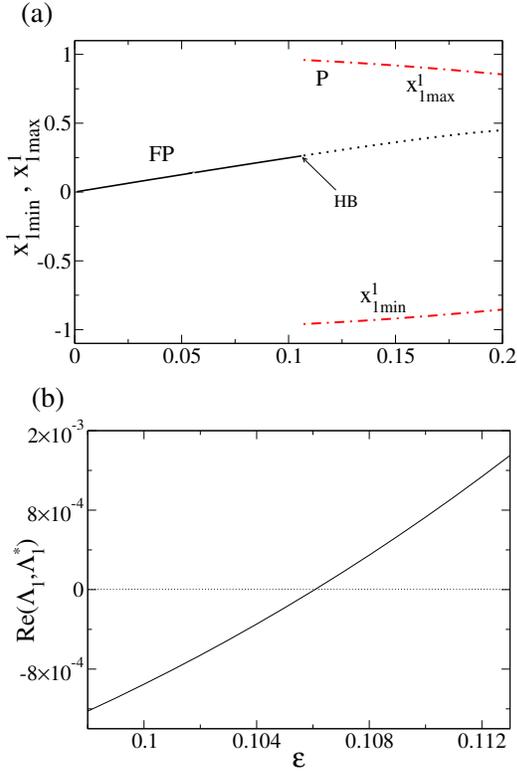

    {\includegraphics[width=0.38\textwidth]{fig04a.eps}\\
    \includegraphics[width=0.37\textwidth]{fig04b.eps}}
  \caption{(Color online)
(a) Bifurcation diagram showing the extrema (red dash-dotted lines) of the variable $x_1^{1}$ [in Eq.~(\ref{eq:3DC3})] as a function of $\varepsilon$ for the values $\varrho=1$, $\mu^{(1)}_{13}=0.3$ and $\mu^{(1)}_{12}=-0.2$. One of the two fixed points of the system, calculated using Eq.~(\ref{eq:Ch3DC4}), is shown in black and marked FP. 
(b) The (equal) real parts of the complex conjugate pair of eigenvalues associated with the fixed points which become positive at $\varepsilon\approx 0.106$ marking a Hopf bifurcation (HB).}
    \label{fig:4} 
\end{figure}    
\end{center}

\begin{center}
\begin{figure}[htp]
{\includegraphics[width=0.35\textwidth]{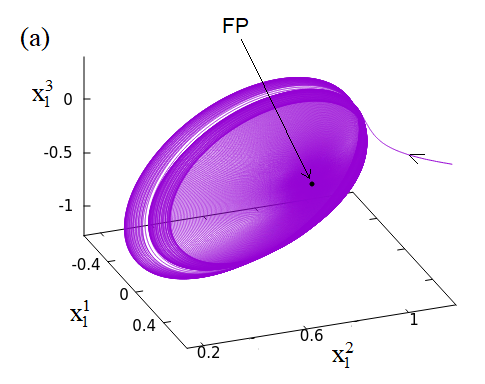}\\ 
 \includegraphics[width=0.37\textwidth]{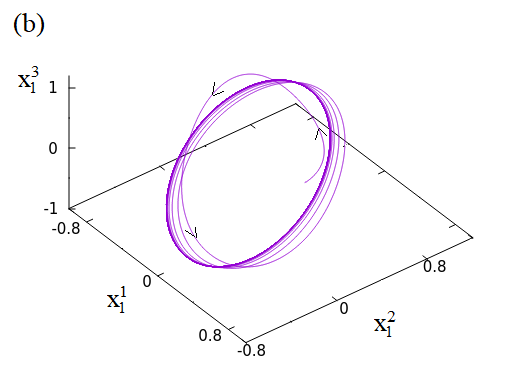}}
 \caption{(Color online) The evolution of $\x_1$, Eq.~(\ref{eq:3DC3}), corresponding to $\gamma=1$, $\mu^{(1)}_{13} = 0.2$, and $\mu^{(1)}_{12} = -0.3$. 
 (a) For $\varepsilon = 0.01$, it gradually approaches the fixed point FP, tracing out a portion of the sphere $r_1\approx\sqrt{\varrho}$ [see Eq.~(\ref{eq:Ch3radial})]. (b) For $\varepsilon = 0.18$, the trajectory approaches a limit cycle.}
\label{fig:5}
\end{figure}
\end{center}

There is a subcritical Hopf bifurcation in the system, as can be seen by fixing, say
$\mu_{12}^{(1)}$ and $\mu_{12}^{(2)}$, while varying $\varepsilon$. When  $\varepsilon = 0$,
the fixed points have coordinates $\pm \sqrt{\varrho} \big(0,\mu^{(1)}_{13},
-\mu^{(1)}_{12}\big)\big{/}\sqrt{(\mu^{(1)}_{12})^2+(\mu^{(1)}_{13})^2}$, which are located 
at the intersections of the axis along $\boldsymbol{\Omega}^{(1)}$ and the sphere of radius 
$\sqrt{\varrho}$. With varying $\varepsilon$, the locations of these fixed points, as well 
as the nature of their stability, change. Shown in Fig.~\ref{fig:4}(a) is the $x^1_1$ component 
of one of the fixed points. At the Hopf bifurcation, this loses stability and a cycle (marked in red)
is born. The precise values of $\varepsilon$ when this occurs depends on the other parameters.
The real parts of the eigenvalues at the fixed point is shown in Fig.~\ref{fig:4}(b), confirming 
Hopf bifurcation. Trajectories of the system are shown in Figs.~\ref{fig:5}(a,b), before and after the
bifurcation point. 

It is also worth highlighting that the limit cycle as well as the fixed points in this case describe solutions that lie on a surface 
that is very close to the sphere $r_1=\sqrt{\varrho}$. The stable 
manifold is defined by ($x^1_1 = x^1_2,\, x^2_1 = x^2_2,\,x^3_1 = - x^3_2$), and  $r_1$ satisfies
[cf.  Eq.~(\ref{eq:3DC3})]
\begin{equation}  
\dot{r}_1 = (\varrho-r_1^2) r_1 - \frac{2\varepsilon(x^{3}_{\small 1})^{2}}{r_1}.
\label{eq:Ch3radial}
\end{equation}
Near the bifurcation, the second term can be treated as a perturbation,
consequently trajectories are attracted towards a sphere defined by $r_1\approx\sqrt{\varrho}$ and 
move along its surface as they approach the attractor; the trajectory 
in Fig.~\ref{fig:5}(a) traces out this surface as it asymptotically reaches the fixed point. 

\begin{center}
\begin{figure}[htp]
     {\includegraphics[width=0.4\textwidth]{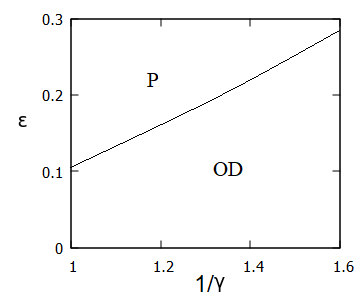}}
     \caption{Schematic phase diagram showing regions of periodicity (P) and oscillation death (OD) as a function of $\gamma$ and $\varepsilon$  for the coupled system of Eq.~(\ref{eq:3DC3}).}
      \label{fig:6}
\end{figure} 
\end{center}

Similar collective behavior is observed even when $\gamma$ deviates from 1, with
only two kinds of dynamics being present: oscillation death and 
limit cycles characterized by phase locking. The $\varepsilon$-$\gamma$ space is thus 
divided into two regions corresponding to periodicity and OD, as shown schematically in 
Fig.~\ref{fig:6}, where the locus of the Hopf bifurcation points have been obtained numerically. 
It may be concluded, therefore, that a misalignment of the axes of the frequency (bi-)vectors of the two oscillators 
leads to the emergence of OD rather than AD (cf. Fig.~\ref{fig:1}), a feature that could be 
relevant for control applications.

\subsection{Coupled oscillators in $D=4$}
The equations of motion for two coupled four-dimensional oscillators with bivectors 
$\boldsymbol{\mu}^{(1)}$ and $\boldsymbol{\mu}^{(2)}$ are
\begin{align}  
\dot{x}^1_n &= (\varrho - r_n^2) x^1_n + \mu^{(n)}_{12} x^2_n + \mu^{(n)}_{13} x^3_n + \mu^{(n)}_{14} x^4_n \nonumber \\
&\qquad\qquad\qquad\qquad\qquad\qquad + \varepsilon (x^1_m - x^1_n), \nonumber\\
\dot{x}^2_n &= (\varrho - r_n^2) x^2_n -\mu^{(n)}_{12} x^1_n + \mu^{(n)}_{23} x^3_n + \mu^{(n)}_{24} x^4_n  \nonumber\\
&\qquad\qquad\qquad\qquad\qquad\qquad + \varepsilon (x^2_m - x^2_n), \nonumber\\
\dot{x}^3_n &= (\varrho - r_n^2) x^3_n -\mu^{(n)}_{13} x^1_n - \mu^{(n)}_{23} x^2_n + \mu^{(n)}_{34} x^4_n \nonumber \\
&\qquad\qquad\qquad\qquad\qquad\qquad + \varepsilon (x^3_m - x^3_n), \nonumber\\
\dot{x}^4_n &= (\varrho - r_n^2) x^4_n - \mu^{(n)}_{14} x^1_n - \mu^{(n)}_{24} x^2_n - \mu^{(n)}_{34} x^3_n\nonumber \\
&\qquad\qquad\qquad\qquad\qquad\qquad + \varepsilon (x^4_m - x^4_n), \label{eq:4D}
\end{align}
where $(n,m)=(1,2)$ or (2,1) as before. In the absence of coupling, the two eigenfrequencies
corresponding to arbitrary values of the parameters $\mu^{(1,2)}_{ij}$ are generically 
incommensurate and result in quasiperiodic dynamics \cite{GGGPR}. Moreover, the 
eigenvectors of the matrices $\mathsf{M}^{(1)}$ and $\mathsf{M}^{(2)}$ corresponding to the bivectors 
$\boldsymbol{\mu}^{(1)}$ and $\boldsymbol{\mu}^{(2)}$ will also not be aligned. Below we 
explore the effect of coupling on the behavior of these two oscillators for various choices of the bivectors.
The difference from the $D =$ 3 case is the absence of a zero-frequency mode. \\

\subsubsection{Common eigenvectors}
When the bivectors are related by $\boldsymbol{\mu}^{(1)} = \gamma\boldsymbol{\mu}^{(2)}$ with 
$\gamma\in(0,1]$, there is a common set of eigen-axes for the two systems and the equations of
motion in normal coordinates assume a simpler form,
\begin{align}
\dot{y}^1_n &= [\varrho - (r_n)^2] y^1_n - \omega^{(n)}_1 y^2_n + \varepsilon (y^1_m - y^1_n), \nonumber\\
\dot{y}^2_n &= [\varrho - (r_n)^2] y^2_n + \omega^{(n)}_1 y^1_n + \varepsilon (y^2_m - y^2_n), \nonumber\\
\dot{y}^3_n &= [\varrho - (r_n)^2] y^3_n - \omega^{(n)}_2 y^4_n + \varepsilon (y^3_m - y^3_n), \nonumber\\
\dot{y}^4_n &= [\varrho - (r_n)^2] y^4_n + \omega^{(n)}_2 y^3_n + \varepsilon (y^4_m - y^4_n),
\label{eq:4D1}
\end{align}
where $(\omega_1^{(1)},\omega_2^{(1)})$ and $(\omega_1^{(1)},\omega_2^{(1)}) = 
\gamma(\omega_1^{(2)}, \omega_2^{(2)})$ 
are the eigenfrequencies corresponding to the two oscillators.

\begin{center}
\begin{figure}[htp]
     {\includegraphics[width=0.4\textwidth]{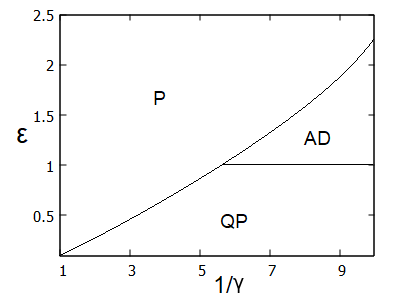}}
     \caption{{\color{black} Schematic presentation of the periodic, quasiperiodic and AD regions constituting the $\gamma-\varepsilon$ parameter space corresponding to Eq.~(\ref{eq:4D1}), drawn for the parameter values $\omega_1^{(1)}=-\sqrt{3+2\sqrt{2}}$ and $\omega_2^{(1)}=-\sqrt{3-2\sqrt{2}}$.}}
      \label{fig:7}
\end{figure} 
\end{center}

\begin{center}
\begin{figure}[htp]
     {\includegraphics[width=0.45\textwidth]{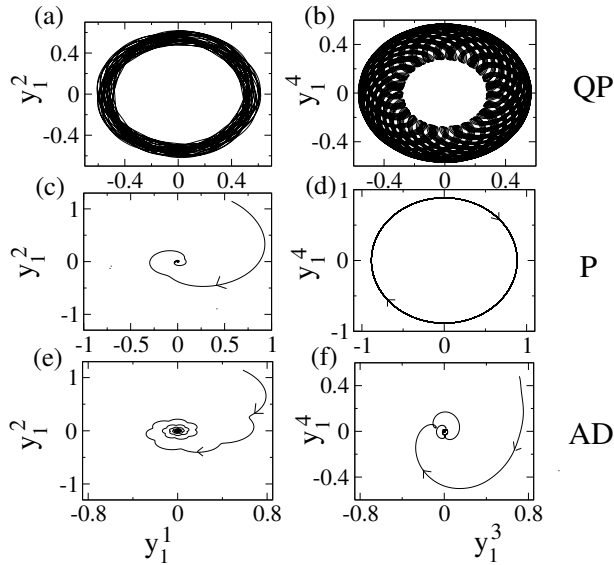}}
     \caption{Quasiperiodic motion in the variable pairs $(y_1^1,y_1^2)$ and $(y_1^3,y_1^4)$ corresponding to Eq.~(\ref{eq:4D1}) for $\varepsilon=0.5$ and $\gamma=0.2$ (QP region in Fig.~\ref{fig:7}) is apparent in (a) and (b). At $\varepsilon=1$ and $\gamma=0.25$ (P region of Fig.~\ref{fig:7}), on the other hand, there is (c) oscillation quenching in the pair $(y_1^1,y_1^2)$ and (d) periodicity in $(y_1^3,y_1^4)$ with the dynamics settling at a limit cycle. Similarly, AD is observed in both the pairs when $\varepsilon=1.2$ and $\gamma=0.125$ (AD region in Fig.~\ref{fig:7}), as shown in (e) and (f).}
      \label{fig:8}
\end{figure} 
\end{center}

Taking, for purposes of illustration, the parameters for one of the oscillators as $\omega_1^{(1)}=-\sqrt{3+2\sqrt{2}}$ 
and $\omega_2^{(1)}=-\sqrt{3-2\sqrt{2}}$ which result from the choice of matrix elements 
$\mathsf{M}^{(1)}_{jk} = 1$ for all $j>k$, we obtain the phase diagram shown in Fig.~\ref{fig:7}; 
the two largest Lyapunov exponents help to characterize the 
regions marked AD, QP and P in the $\gamma$-$\varepsilon$ plane. 
In the periodic domain, a subset of variables corresponding to each oscillator 
undergo oscillation quenching, while the rest oscillate periodically, {\color{black}describing a special form of 
{\em partial amplitude death} \cite{Liu}. Figure~\ref{fig:7} describes interesting transitions from AD and QP to these unique partial AD states.} 
Orbits are shown in Figs.~\ref{fig:8}(a) and \ref{fig:8}(b) where the coordinate pairs $(y_1^1,y_1^2)$ 
and $(y_1^3,y_1^4)$ are plotted for $\varepsilon = 0.5$, ${\gamma} = 0.2$ in the quasiperiodic regime. For $\varepsilon = 1$ and $\gamma=0.25$, oscillation quenching in $(y_1^1,y_1^2)$ and limit cycle oscillations in $(y_1^3,y_1^4)$ are shown in 
Figs.~\ref{fig:8}(c) and (d). Amplitude death occurs for parameters $\gamma = 0.125$ 
and $\varepsilon = 1.2$;  see Figs.~\ref{fig:8}(e) and (f). 

In the partial amplitude death observed in studies of ensembles of coupled dynamical systems, the 
variables of individual elements are driven to fixed points \cite{Umeshkanta,Liu}. As can be seen 
easily from Eq.~(\ref{eq:4D1}),  the four-dimensional subspace with coordinates 
$(y_1^1,y_1^2,y_2^1,y_2^2) = (0,0,0,0)$ is invariant: initial points with  $\left(0,0,y_1^3(0),y_1^4(0)\right)$ 
and $\left(0,0,y_2^3(0),y_2^4(0)\right)$ respectively for the two oscillators, stay in the subspace for all times 
$t > 0$.  The case of partial amplitude death therefore occurs when the coupling stabilizes this subspace,
and this can happen in the region marked P in Fig.~\ref{fig:7}. The similarity to Fig.~\ref{fig:1} is not 
a coincidence: partial amplitude death effectively reduces the system to a pair of coupled oscillators in 
$D = 2$. This implies that the radii of the limit cycles as well as the relative phases for 
the phase-locked states corresponding to region P of Fig.~\ref{fig:7} can be obtained analytically 
from earlier results \cite{Aronson}. They are $r_*^2 = 1 -\varepsilon + \sqrt{\varepsilon^2 - 
\Delta^2/4}$ and $\alpha_*=\sin^{-1}(\Delta/2\varepsilon)$, 
where $\Delta = \omega_2^{(1)} \big(1-\gamma^{-1}\big)$. It is also straightforward 
to show that quasiperiodicity occurs when $|\Delta| < 2$ and $\varepsilon\in \big(0,\tfrac{1}{2} |\Delta|\big)$, 
or when $|\Delta|>2$ and $\varepsilon\in (0,1)$; AD corresponds to $|\Delta|>2$ and 
$\varepsilon\in \left(1,\tfrac{1}{2} \big(1+\tfrac{1}{4}\Delta^2)\right)$; and periodicity occurs when 
$|\Delta| < 2$ and $\varepsilon > \tfrac{1}{2} |\Delta|$ or when $|\Delta|>2$ and 
$\varepsilon > \tfrac{1}{2} \big(1 + \tfrac{1}{4} \Delta^2\big)$. 
For other sets of incommensurate frequencies $(\omega_1^{(1)},\omega_2^{(1)})$, the characteristic 
phase diagrams are similar to Fig.~\ref{fig:7}, with the transition points being determined by $|\Delta|$.

The $D =$ 4 oscillators can be brought into quaternionic form with $\omega_1^{(1)} = \omega_2^{(1)} = \omega^{(1)}$ 
and $\omega_1^{(2)} = \omega_2^{(2)} = \omega^{(2)}$. If we now choose $\omega^{(1)} = \gamma\omega^{(2)}$, 
and analyze the collective dynamics, the $\gamma$-$\varepsilon$ space again separates into three 
different dynamical regions: QP, P, and AD, similar to the case of two coupled two-dimensional SL oscillators.\\

\subsubsection{Distinct eigenvectors}
This is the more general case since whenever the two 
bivectors are not related by scaling, namely if $\boldsymbol{\mu}^{(1)} \neq \gamma\boldsymbol{\mu}^{(2)}$, 
their eigenaxes will be distinct. We have made only a limited exploration of the general case and our
observation is that depending on the choice of the bivectors and the coupling strength $\varepsilon$, 
there can be AD, OD, quasiperiodicity and periodicity. Given the large number of parameters in the 
components of the bivectors,  there can be a variety of transitions between the different dynamical states. 

\begin{center}
    \begin{figure}[htp]
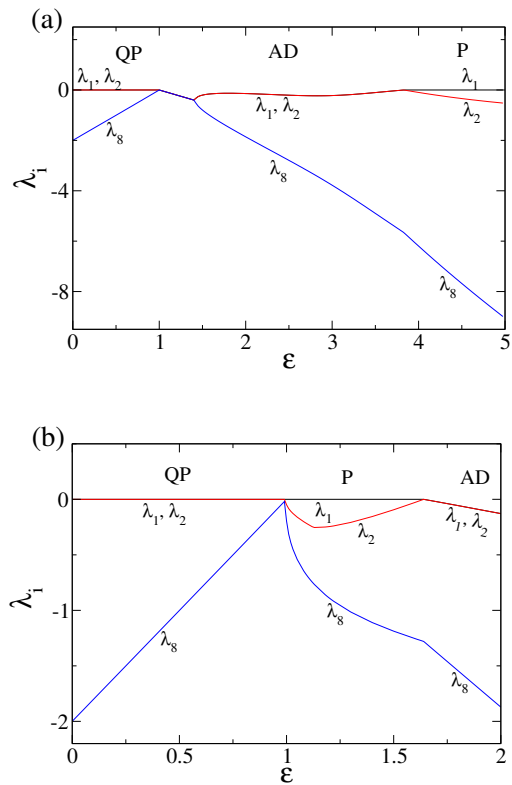

     \includegraphics[width=0.37\textwidth]{fig09a.eps}\\
     \vspace{7mm}\includegraphics[width=0.37\textwidth]{fig09b.eps}
     \caption{(Color online) 
     The bivector  $\boldsymbol{\mu}$ can be specified by its six independent components: $\{ \mu_{12}, \mu_{13},\mu_{14}; \mu_{23}, 
     \mu_{24}; \mu_{34} \}$.
     In (a) we take $\boldsymbol{\mu}^{(1)} = (0.6, 0.2, 2.1; 
     1.0, 0.5; 0.2)$ and $\boldsymbol{\mu}^{(2)} = 10 (0.2, 2.1, 1.0; 0.2, 0.6; 0.5)$. The spectrum of the 
     Lyapunov exponents $\lambda_1$, $\lambda_2$ and $\lambda_8$ is plotted as a function of $\varepsilon$. 
     Q, AD, and P denote quasiperiodicity, amplitude death and periodic dynamics. (b) A transition from 
     quasiperiodicity to periodicity to AD, when the second bivector is quaternionically reduced. The 
     spectrum of eight Lyapunov exponents  is shown for the case $\boldsymbol{\mu}^{(1)} = (10, 2, 3; 
     0, 0; 3)$ and $\boldsymbol{\mu}^{(2)} = (-1, -2, -3; -3, 2; -1)$.}
     \label{fig:9} 
\end{figure}
\end{center}

For illustrative purposes, we show the results for the choice $|\boldsymbol{\mu}^{(1)}| = \gamma
|\boldsymbol{\mu}^{(2)}|$, namely by randomly permuting the components of $\boldsymbol{\mu}^{(1)}$ and 
scaling them by an overall factor in order to generate $\boldsymbol{\mu}^{(2)}$. For a 
representative case, the results are summarized in Fig.~\ref{fig:9}, where the Lyapunov spectrum
is plotted as a function of the coupling (the components of the bivectors are in the caption). 

A transition from {\em quasiperiodic} motion to {\em periodic} dynamics is followed by {\em amplitude 
death} in Fig.~\ref{fig:9}(a) where the exponents $\lambda_1$, $\lambda_2$, and $\lambda_8$ are shown 
as a function of $\varepsilon$. In other ranges of the parameters, a transition from {\em quasiperiodicity} 
to {\em oscillation death} to {\em periodicity} may be observed. A transition from {\em quasiperiodicity} to 
{\em periodicity} to {\em amplitude death}, shown in Fig.~\ref{fig:9}(b), can be seen in a related example, 
where one of the oscillators has its parameters reduced to be quaternionic. In this case one quasiperiodic 
oscillator is coupled to a limit cycle. In both Figs. \ref{fig:9}(a) and  \ref{fig:9}(b), the QP regions are characterized by 
phase drift, while the P regions involve phase synchronization between the oscillators. 

\section{Symmetry-Breaking Coupling}

In the present formalism, since the rotational symmetries are explicitly built into 
the structure of the generalized Stuart-Landau equations, the role of symmetry can 
be distinguished from the role of the coupling term since rotational symmetries can be 
selectively broken. Note though that the $D$-dimensional Stuart Landau equations 
have the non-Abelian SO($D$) symmetry, therefore, only specific combinations 
of the symmetries can be broken. 

We focus here on dynamics that results as a {\em consequence} of symmetry-breaking, and therefore
discuss only those aspects that are not seen in couplings that preserve the symmetry. Of 
necessity, only a limited exploration of symmetry-breaking is possible since the number of 
parameters is large and it is possible to choose many different forms of coupling. 
We look at some representative cases, retaining the linear diffusive coupling 
between the oscillators and breaking symmetries by adjusting the coupling. Rewriting 
Eq.~(\ref{eq:3D2a}) to give additional flexibility on the coupling constants, we have the 
equation for the coupled oscillators as 
\begin{align}
&\dot{y}^1_n = [\varrho - (r_n)^2] y^1_n - \Omega^{(n)} y^2_n + \varepsilon_n^1 (y^1_m - y^1_n), \nonumber\\
&\dot{y}^2_n = [\varrho - (r_n)^2] y^2_n + \Omega^{(n)} y^1_n + \varepsilon_n^2 (y^2_m - y^2_n), \nonumber\\
&\dot{y}^3_n = [\varrho - (r_n)^2] y^3_n + \varepsilon_n^3 (y^3_m - y^3_n), \label{eq:3D2}
\end{align}

where $\Omega^{(1)}=\gamma\Omega^{(2)}$, and $(n,m) = (1,2)$ or $(2,1)$.

\begin{center}
     \begin{figure}[htp]
     \includegraphics[width=0.32\textwidth]{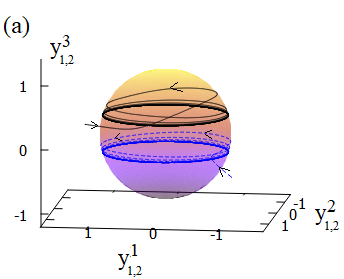}\\
     \vspace{3.4cm}\includegraphics[width=0.5\textwidth]{fig10b.eps}
     \caption{(Color online) Trajectories (solid black and dashed blue lines) of the two oscillators 
     coupled through the pair $(y^2_1, y_2^2)$ in Eq.~(\ref{eq:3D2}) may reach (a) two different limit 
     cycles, with the oscillators being partially synchronized with complete alignment in the pairs 
     $(y_1^i, y_2^i)$, $i=1,2$, as shown in (b) for $i=1$. Plot (c) shows {\em partial oscillation death} 
     in which the constant asymptotic values of $y_1^3$ (solid black) and $y_2^3$ (dotted red) differ in 
     sign. (Parameters for the plots: $\gamma=1$, $\varepsilon=0.5$, and $\Omega^{(1)}=0.2$).}
    \label{fig:10}
    \end{figure}
\end{center}

For instance, when the frequences of the oscillators are identical, namely  $\Omega^{(1)}=\Omega^{(2)}$, 
and $\varepsilon_1^2=\varepsilon_2^2=\varepsilon>0$, the other $\varepsilon_n^i$'s being 0, they are only 
coupled via $(y_1^2,y_2^2)$. Coupling of oscillators via a single variable pair is in fact well explored 
in the literature \cite{Peco1,Kurths}. In contrast to the symmetry-preserving coupling, here we find that 
for initial conditions  $(y_1^3(0)>0,y_2^3(0)<0)$ or $(y_1^3(0)<0,y_2^3(0)>0)$, the oscillators 
approach two inhomogeneous limit cycles \cite{Prag,Tanmoy1} shown in Fig.~\ref{fig:10}(a). There are 
alignments between the pairs $y_1^1=y_2^1$ and $y_1^2=y_2^2$, as a result of the coupling
(see Appendix.~\ref{Ap:6}), as shown in Fig.~\ref{fig:10}(b), and additionally $y_1^3=-y_2^3$, as shown 
in Fig.~\ref{fig:10}(c). Since the variables $y_1^3$ and $y_2^3$ here asymptote to two different values, 
we label this phenomenon {\em partial oscillation death}, to distinguish it from partial amplitude death 
reported earlier \cite{Umeshkanta, Liu}. For other initial conditions, on the other hand, there is a complete 
synchronization between the oscillators (not shown). Thus a different dynamical state of coexistence of partial and complete synchronization can be seen to result from the symmetry-breaking coupling. The dynamics is furthermore characterized by extreme multistability. 

\begin{figure}[htp]
      \centering
      \begin{tabular}{cc}
    \hspace{-0.5cm} \includegraphics[width=0.25\textwidth]{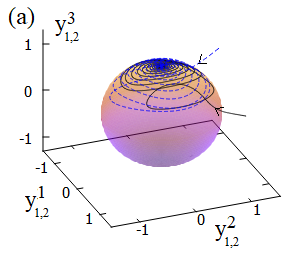}
&
    \includegraphics[width=0.22\textwidth]{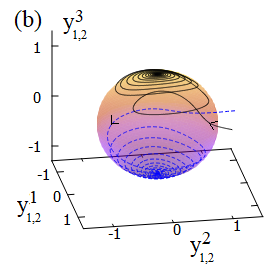}\\
      
    \hspace{-0.5cm}  \includegraphics[width=0.25\textwidth]{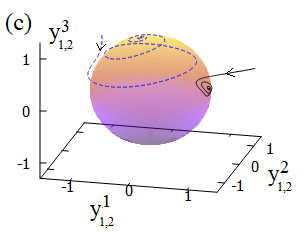}
 &
      \includegraphics[width=0.25\textwidth]{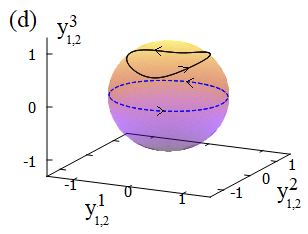}
   \end{tabular}
     \caption{(Color online) Depending on the initial conditions, trajectories of the two oscillators in Eq.~(\ref{eq:3D2}), coupled via the pair $(y_1^2,y_2^2)$, approach one of the four stable fixed points, (a) $(0,0,\sqrt{\varrho};0,0,\sqrt{\varrho})$, (b) $(0,0,\sqrt{\varrho};0,0,-\sqrt{\varrho})$, plotted for $\gamma=0.5$, $\varepsilon=0.7$. When one of the oscillators starts on the equatorial plane, the trajectories may reach (c) two fixed points, plotted for $\gamma=1/6$ and $\varepsilon=0.7$ or (d) two different limit cycles, plotted for $\gamma=1/2$ and $\varepsilon=0.7$. Parameter $\Omega^{(1)}=0.2$ is used for all the plots.}
      \label{fig:11}
\end{figure}

When the oscillators are not identical but have a common axis of rotation, symmetry-breaking 
can lead to coexisting AD and OD states (not observed in symmetry-preserving coupling). 
Consider the coupling type of the previous example, namely $\varepsilon_1^2=\varepsilon_2^2=\varepsilon>0$, 
but for two {\em nonidentical} oscillators whose characteristic frequencies are related by $\Omega^{(1)}=
\gamma\Omega^{(2)}$, where $\gamma\in(0,1)$; see Eq.~(\ref{eq:3D2}). For initial conditions satisfying 
$\left(y_1^3(0)>0,y_2^3(0)>0\right)$, or $\left(y_1^3(0)<0,y_2^3(0)<0\right)$; the two systems reach the stable fixed points 
$(0,0,\sqrt{\varrho}; 0,0,\sqrt{\varrho})$, or $(0,0,-\sqrt{\varrho}; 0,0,-\sqrt{\varrho})$, respectively, 
as depicted in Fig.~\ref{fig:11}(a), signifying AD. However, for another class of initial points 
$\left(y_1^3(0)>0,y_2^3(0)<0\right)$ [or $(y_1^3(0)<0,y_2^3(0)>0)$] the two oscillators reach the fixed points 
$(0,0,\sqrt{\varrho}; 0,0,-\sqrt{\varrho})$, or $(0,0,-\sqrt{\varrho}; 0,0,\sqrt{\varrho})$ [as depicted 
in Fig.~\ref{fig:11}(b)], signifying OD. When only one of the oscillators starts from the equatorial plane, there is a possibility of a different OD state, where one oscillator settles at a fixed 
point on the equatorial plane while the other one settles near one of the poles, as depicted in 
Fig.~\ref{fig:11}(c), as well as the possibility of a periodic state, as shown in Fig.~\ref{fig:11}(d). 
When both the oscillators start from the equatorial plane, the coupled system effectively reduces to a
system of two two-dimensional SL oscillators coupled in a single variable pair, a case studied earlier 
by Koseska {\em et al.} \cite{Kurths}. Note that when the coupling only involves the pair $(y_1^2,y_2^2)$ 
in Eq.~(\ref{eq:3D2}), the subspaces $y_1^3=0$ and $y_2^3=0$ become invariant manifolds and consequently 
the trajectories never cross the equatorial plane, as in the cases depicted in Figs.~\ref{fig:10} 
and ~\ref{fig:11}.

Similarly, when the coupling between two identical oscillators of Eq.~(\ref{eq:3D2}) only involves 
the pair $(y_1^3,y_2^3)$ (that is $\varepsilon_1^3=\varepsilon_2^3=\varepsilon>0$ while the others
are zero), there is partial synchronization in the pair $y_1^3=y_2^3$, while the rest of the pairs 
are phase synchronized with a relative phase determined by the initial condition (not shown). On 
the other hand, when the two oscillators are not identical, partial synchronization in the pair 
$y_1^3=y_2^3$ is accompanied by phase-drift between the remaining pairs (not shown).

For two three-dimensional oscillators with axes of rotation that are not parallel, while 
symmetry-preserving coupling can only lead to phase synchronization (see the P region in Fig.~\ref{fig:6}), 
symmetry-breaking couplings can lead to {\em partial synchronization} between the periodic states. This 
occurs when the two oscillators, which evolve by $\dot{\x}_{n} = (\varrho-r_n^2)\x_n + 
\boldsymbol{\Omega}^{(n)}\times\x_n + \varepsilon(x_m^1-x_n^1)\e_1$, where $(n,m)=(1,2)$ or $(2,1)$, 
have the same characteristic frequency $\Omega^{(1)}=\Omega^{(2)}$. The asymptotic states here 
obey the equality $x_1^1=x_2^1$, while the rest of the variable pairs show phase synchronization. On the 
other hand, when $\Omega^{(1)} \neq \Omega^{(2)}$, coupling stabilizes the fixed points lying at the 
intersections of the two axes $\hat{\boldsymbol{\Omega}}^{(1)}$ and $\hat{\boldsymbol{\Omega}}^{(2)}$ 
with the sphere $\mathbb{S}^{(2)}$ of radius $\sqrt{\varrho}$, leading to AD.

Similarly, whereas symmetry-preserving coupling only allows completely synchronized quasiperiodic states, 
symmetry-breaking leads to partially synchronized inhomogeneous quasiperiodic states for a coupled 
$D=4$ system described by 
\begin{align}
\dot{y}^1_n &= (\varrho - r_{n}^2) y^1_n - \omega^{(n)}_1 y^2_n + \varepsilon_n^1 (y^1_m - y^1_n),\nonumber\\
\dot{y}^2_n &= (\varrho - r_n^2) y^2_n + \omega^{(n)}_1 y^1_n + \varepsilon_n^2 (y^2_m - y^2_n), \nonumber \\
\dot{y}^3_n &= (\varrho - r_n^2) y^3_n - \omega^{(n)}_2 y^4_n + \varepsilon_n^3 (y^3_m - y^3_n),  \nonumber\\
\dot{y}^4_n &= (\varrho - r_n^2) y^4_n + \omega^{(n)}_2 y^3_n + \varepsilon_n^4 (y^4_m - y^4_n).
\label{eq:4D2}
\end{align}
\vspace{3.6cm}

\begin{center}
   \begin{figure}[htp]
     \includegraphics[width=0.47\textwidth]{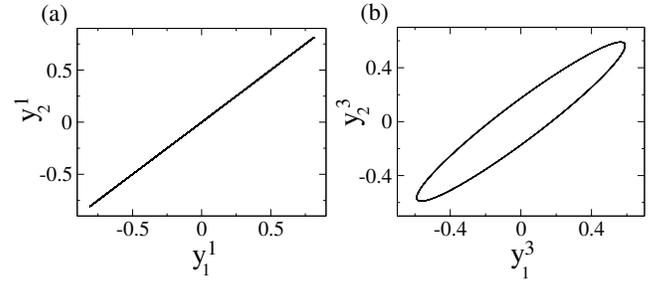}
     \caption{For two identical oscillators of Eq.~(\ref{eq:4D2}) coupled symmetrically in the pair 
     $(y_1^1,y_2^1)$, (a) complete alignment of variables $y_1^1=y_2^1$ and $y_1^2=y_2^2$, and (b) 
     the remaining pairs show phase synchrony. (Plotted for $\varepsilon_1^1=\varepsilon_2^1=1$.)}
     \label{fig:12} 
   \end{figure}
\end{center}

When the frequencies of the oscillators are 
identical, i.e., $\omega^{(1)}_1=\omega^{(2)}_1$ and $\omega^{(1)}_2=\omega^{(2)}_2$, and the coupling only 
involves the pair $(y_1^1,y_2^1)$, a coexistence of complete synchronization and partial synchronization 
involving the pairs $y_1^1=y_2^1$ and $y_1^2=y_2^2$, as shown in Fig.~\ref{fig:12} is observed (see Appendix
\ref{Ap:6} for an explanation). In case of partial synchronization, the two oscillators approach two 
{\em different} quasiperiodic attractors determined by their initial conditions, both of which are located 
on the sphere $\mathbb{S}^3$ of radius $\sqrt{\varrho}$. Interestingly, when the oscillators are nonidentical 
with $\omega_1^{(1)}=\gamma\omega^{(2)}_1$ and $\omega_2^{(1)}=\gamma\omega^{(2)}_2$, $\gamma\in(0,1)$; the 
coupling induces partial amplitude death in the variables $y_1^1=y_2^1=y_1^2=y_2^2=0$ in the asymptotic limit. 
This results in periodic dynamics of the two oscillators characterized by phase drift, whereas symmetry-preserving 
coupling, only allows phase synchronization between the periodic states (refer to the P region in Fig.~\ref{fig:7}). 
Similarly with coupling in the pair $(y_1^3,y_2^3)$, partial synchrony is observed between the pairs 
$y_1^3=y_2^3$ and $y_1^4=y_2^4$ when the oscillators are identical, and partial amplitude death with 
$y_1^3=y_2^3=y_1^4=y_2^4=0$ is observed when the oscillators are nonidentical. 

Another interesting scenario arises when two quaternionically reduced nonidentical oscillators with different 
eigenaxes are coupled symmetrically in the coordinate pair $(x_1^1,x_2^1)$, which are not eigen-coordinates. 
In this case, partial AD in the pair $x_1^1=x_2^1=0$ is observed, with the two systems settling at two 
three-dimensional inhomogeneous limit cycles, each outlining a great circle (not shown). On the other hand, 
when the bivectors for the reduced oscillators satisfy $|\boldsymbol{\mu}^{(1)}|=|\boldsymbol{\mu}^{(2)}|$, 
partial synchronization in the pair $x_1^1=x_2^1$ is observed (see Appendix \ref{Ap:6}) for an explanation.

The symmetry-breaking cases discussed so far only involve partial breaking of symmetry. In each 
of these cases the coupled system retains its form under a subgroup rotations, namely those not involving 
selected by the (nonvanishing) couplings. In contrast, if, for example, $\varepsilon_1^1=\varepsilon_2^1$, 
$\varepsilon_1^2=\varepsilon_2^2$, but $\varepsilon_1^1\neq\varepsilon_1^2$ in Eqs.~(\ref{eq:3D2}) and 
(\ref{eq:4D2}), symmetry is broken {\em completely}. With this type of coupling, we have observed complete 
synchronization and AD (not shown).


\section{Summary and Discussion}

In this paper we have studied the collective behavior of coupled Stuart-Landau oscillators in $D=3$ and 
$D=4$ dimensions, generalizing the earlier work in $D=2$ by Aronson {\em et al} \cite{Aronson}. Three and 
four dimensions are representative  of odd and even dimensions, in which the system behaves slightly 
differently. We have also restricted our study mostly to two coupled oscillators, since it has not been studied in
higher dimensional generalizations of the Stuart-Landau system. The scope of this study is vast: in 
addition to the parameters of each individual oscillator, there are coupling constants and forms of 
couplings that can be varied. As a consequence we have been able to explore only a limited part of the 
parameter space and have examined the most commonly studied linear diffusive coupling, for both  
symmetry-preserving and symmetry-breaking cases. The contrasting results, which we tabulate in 
Tables~\ref{T1} and \ref{T2}, highlight the importance of symmetry.

Needless to say, many other forms of coupling that preserve or break rotational 
symmetry \textcolor{black}{(naturally, of the normal form)} can 
be considered, analysis of which is a rich and promising avenue for future research.
For instance, nonlinear coupling of the form $\varepsilon_{1,2}(|\x_{2,1}|^2\x_{2,1}-|\x_{1,2}|^2\x_{1,2})$ 
preserve rotational symmetry; however, the dynamics is expected to be very different. Similarly 
for broken symmetry case, conjugate coupling of Ref.~\cite{Nirmal} will be a natural addition to 
the couplings considered here. Likewise, larger networks of SL oscillators with various forms 
of coupling may profitably be studied, the current one being a first step and a guide for further 
exploration. There is potential for novel collective phenomena as the dimension $D$ and the number 
of oscillators $K$ increase. The effect of higher-order interactions in such ensembles is another 
possible extension, an important one considering recent advances \cite{Amico}.

\noindent{\bf Acknowledgements:}
P.B.G., presently at the Department of Dynamics, {\L}{\'o}d{\'z} University of Technology, {\L}{\'o}d{\'z}, acknowledges financial support from the University of Delhi through the University Research 
Fellowship. R.R. thanks the Department of Physics, IIT-Delhi, for hospitality during the time this work was initiated. We thank Rahul Ghosh for discussions.

{\scriptsize
\begin{table*}
\caption{Two coupled three-dimensional SL oscillators}
\begin{center}
\setlength{\tabcolsep}{6pt}
{\renewcommand{\arraystretch}{1.8}
\begin{tabular}{| p{0.13\textwidth} | p{0.4\textwidth} | p{0.4\textwidth} |}
\hline
{} & {\bf Symmetry-preserving coupling} & {\bf Symmetry-breaking coupling}\\

\hline

Case I.  

$\boldsymbol{\Omega}^{(2)} = \boldsymbol{\Omega}^{(1)}$

& 
Completely synchronized periodic states.

& 

Periodic (phase synchronization) states;

coexistence of (a) periodic (complete synchronization) and 
(b) {\em periodic (partial synchronization) states characterized by partial OD} (Figs.~\ref{fig:10}).

\\

\hline

Case II.  

$\boldsymbol{\Omega}^{(2)} = \boldsymbol{\Omega}^{(1)}/\gamma$,

$\gamma \in (0,1)$

&
Coexistence of (a) periodic (phase synchronization), (b) AD, and (c) quasiperiodic (phase-drift) states (Figs.~\ref{fig:1} and \ref{fig:2}).

&

{\em Coexistence of (a) periodic (phase synchronization), (b) AD, and (c) OD states} (Fig.~\ref{fig:11});
{\em periodic (phase drift) states}.\\

\hline

Case III.  

$\hat{\boldsymbol{\Omega}}^{(2)}\neq\hat{\boldsymbol{\Omega}}^{(1)}$

&
(a) Periodic (phase synchronization) and (b) OD states (Figs. \ref{fig:5} and \ref{fig:6}). 

& 
{\em Periodic (partial synchronization) states when $|\boldsymbol{\Omega}^{(1)}|=|\boldsymbol{\Omega}^{(2)}|$}, and AD states when $|\boldsymbol{\Omega}^{(1)}|\neq|\boldsymbol{\Omega}^{(2)}|$.\\

\hline
\hline

\end{tabular}}

\end{center}
\label{T1}

\end{table*}
}

{\scriptsize
\begin{table*}
\caption{Two coupled four-dimensional SL oscillators}
\begin{center}
\setlength{\tabcolsep}{6pt}
{\renewcommand{\arraystretch}{1.8}
\begin{tabular}{| p{0.13\textwidth} | p{0.4\textwidth} | p{0.4\textwidth} |}
\hline
{} & {\bf Symmetry-preserving coupling} & {\bf Symmetry-breaking coupling}\\

\hline

Case I.  

$\boldsymbol{\mu}^{(2)} = \boldsymbol{\mu}^{(1)}$

&
Completely synchronized quasiperiodic states.

& 

Coexistence of (a) quasiperiodic (complete synchronization) states, and (b)
{\em inhomogeneous quasiperiodic (partial synchronization) states} (Fig.~\ref{fig:12}).\\

\hline

Case II.  

$\boldsymbol{\mu}^{(2)} = \boldsymbol{\mu}^{(1)}/\gamma$,

$\gamma \in (0,1)$

&
(a) Periodic (phase synchronized) states characterized by partial AD,
(b) quasiperiodic (phase drift) states, and
(c) AD states (Figs.~\ref{fig:7} and ~\ref{fig:8}).

&

{\em Periodic (phase drift) states characterized by partial AD}.\\
\hline
Case III.  

$\boldsymbol{\mu}^{(2)} \neq \boldsymbol{\mu}^{(1)}$

(different eigenaxes)
&
(a) Periodic (phase synchronization), (b) AD, (c) quasiperiodic (phase drift) (Fig. \ref{fig:9}) and (d) OD states. 

& 

(a) Periodic (phase drift) states characterized by partial AD when $|\boldsymbol{\mu}^{(1)}|\neq|\boldsymbol{\mu}^{(2)}|$, and 
(b) {\em periodic (partial synchronization) states when $|\boldsymbol{\mu}^{(1)}|=|\boldsymbol{\mu}^{(2)}|$} (for two oscillators that are reduced quaternionically).\\

\hline\hline
\end{tabular}}
\end{center}
\label{T2}
\end{table*}
}

\appendix
{\color{black}
\section{\label{Ap:1}STABILITY OF THE FIXED POINTS}
To analyze the details of the non-trivial fixed point $(r_{1*},r_{2*},1)$ of Eq.~\eqref{eq:MagAngl} for $\boldsymbol{\mu}^{(1)}=\boldsymbol{\mu}^{(2)}$, we first set the RHS of the equation to zero, and find the relations,
\begin{align}
\varepsilon_2(\varrho_{1*}-r_{1*}^2)r_{1*}+\varepsilon_1(\varrho_2-r_{2*}^2)r_{2*}&=0, \nonumber\\
\frac{r_{1*}}{\varepsilon_1}(r_{1*}^2-\varrho_1)+r_{1*}&=r_{2*}.
\label{eq:A1}
\end{align}
For any given set of $(\varrho_1,\varrho_2,\varepsilon_1,\varepsilon_2)$, the corresponding values of $(r_{1*},r_{2*})$ follow from Eq.~(\ref{eq:A1}). Similarly using the relation
${r_{1*}(r_{1*}^2 + \varepsilon_1 - \varrho_1)}/({\varepsilon_1 r_{2*}})  = 
{r_{2*}(r_{2*}^2 + \varepsilon_2 - \varrho_2)}/({\varepsilon_2 r_{1*}})=1$,
which also follows from Eq.~(\ref{eq:A1}), we obtain the secular determinant for the eigenvalues of the Jacobian corresponding to the fixed point, leading to the cubic equation
\begin{align}
&\left(\lambda + \frac{2f_2}{r_{1*} r_{2*}}\right) \bigg[\lambda^2 
+ \bigg(2(r_{1*}^2 + r_{2*}^2) \nonumber\\ 
&+ \frac{ f_2}{r_{1*} r_{2*}}\bigg]\lambda + 4 r_{1*}^2 r_{2*}^2 
+ \frac{2 f_4}{r_{1*} r_{2*}}\bigg)= 0,
\label{eq:FPStability}
\end{align}
where $f_2 = \varepsilon_2 r_{1*}^2 + \varepsilon_1 r_{2*}^2$ and $f_4 = \varepsilon_2 
r_{1*}^4 + \varepsilon_1 r_{2*}^4$. It is apparent that one of the eigenvalues, $\lambda_1=-\frac{2f_2}{r_{1*} r_{2*}}<0$, while the rest $\lambda_{2,3}=\frac{-b\pm\sqrt{b^2-4ac}}{2a}$, where $a=1,b= 2(r_{1*}^2 + r_{2*}^2)+ \frac{ f_2}{r_{1*} r_{2*}}$ and $c=4 r_{1*}^2 r_{2*}^2 
+ \frac{2 f_4}{r_{1*} r_{2*}}$. When $\varepsilon_{1,2}>0$, $a,b,c>0$, and consequently $\text{Re}(\lambda_{2,3})<0$. This establishes the fixed points $(r_{1*},r_{2*},1)$ are stable, implying the trajectories of the two oscillators approach two spheres of radii $\sqrt{r_{1*}}$ and $\sqrt{r_{2*}}$, while attaining phase synchronization [$\displaystyle{\lim_{t\to\infty}\alpha(t)}=0$], as in Fig.~\ref{fig:3}. As a special case, if $\varrho_1 = \varrho_2 = \varrho$ (in addition to the $\boldsymbol{\mu}^{(1)} = \boldsymbol{\mu}^{(2)}$ condition that has already been assumed), i.e., for a pair of identical Stuart-Landau oscillators that 
are, however, coupled asymmetrically. The fixed point locus is at $(r_{1*}, r_{2*}, \cos\alpha_*) 
= (\sqrt{\varrho}, \sqrt{\varrho}, 1)$, and the two coupled oscillators of 
Eq.~(\ref{eq:MagAngl}) in this scenario approach a sphere of radius $\sqrt{\varrho}$ and are 
completely synchronized in the asymptotic limit, as also pointed out in the main text.

\section{ON THE ANALYTICAL SOLUTIONS OF EQ.~(\ref{eq:MagAngl})}\label{Ap:3}
For $\varrho_1=\varrho_2$, Eq.~(\ref{eq:MagAngl}) is symmetric under the exchange $r_1\leftrightarrow r_2$, 
which means that a solution with $\displaystyle{\lim_{t\to \infty}} r_1, r_2 = r$ may 
exist. In $D=2$, such a solution can be shown to obey \cite{Aronson}:
\begin{equation}
r^2(t)=a(1-c^2)/\Big[1+c\sin(\alpha(t)+\psi)\Big],
\label{eq:2D}
\end{equation}
where $a=1-\varepsilon$, $c={\varepsilon}\big{/}{\sqrt{a^2+\tfrac{1}{4}\Delta^2}}$ and $\tan\psi 
= 2a/\Delta$. For small enough $\varepsilon$, Eq.~(\ref{eq:2D}) describes an ellipse in 
$(r^2,\alpha)$ space, which is stable for appropriate values of $\varepsilon$ and $\Delta$ \cite{Aronson}. 
One concludes that the two-dimensional projections of the flow over $(x_1^1, x^2_1)$ and 
$(x_1^1, x^2_1)$ spaces are annular regions bounded by $r_\pm^2 = a(1 \pm c)$. On the other hand, for 
certain ranges of the parameters $\varepsilon$ and $\Delta$, the presence of a node at 
$r_*^2 = 1 -\varepsilon + \sqrt{\varepsilon^2- \tfrac{1}{4}\Delta^2}$ and 
$\alpha_* = \sin^{-1}\tfrac{\Delta}{2\varepsilon}$ implies that the motion of the individual 
oscillators correspond to limit cycles of radii $r_*$ and phase locking between the two at a 
relative angle $\alpha_*$. These results assume particular significance, since many of the cases of 
coupled higher dimensional oscillators discussed in this paper reduce effectively to two coupled {\em 
two-dimensional} oscillators for certain regimes of parameters.

\section{COUPLING DEPENDENT RADII IN EQ.~(\ref{eq:CpldOsc})}\label{Ap:4}
Let $\rho = {r_{2*}}/{r_{1*}}$ be the ratio of the fixed points $r_{1*}$ and $r_{2*}$ of 
Eq.~\eqref{eq:MagAngl}, for $\boldsymbol{\mu}^{(1)} = \boldsymbol{\mu}^{(2)}$. It can be shown to 
satisfy the algebraic equation   

\textcolor{black}{
\begin{equation}
\rho^4 + \left( \frac{\varrho_1}{\varepsilon_1} - 1\right) \rho^3 - \left(\frac{\varrho_2}{\varepsilon_1} - \frac{\varepsilon_2}{\varepsilon_1}\right) \rho - \frac{\varepsilon_2}{\varepsilon_1} = 0.
\label{eq:ratio}
\end{equation}
}

Since we are interested in the case of very strong coupling, we set $\varepsilon_1 = \varepsilon_2 = 
\varepsilon$ and consider the limit $\varepsilon \to \infty$. Then Eq.~(\ref{eq:ratio}) simplifies 
to $\rho^4 - \rho^3 + \rho - 1 = 0$ for which $\rho = 1$ is a solution, implying that the limiting 
spheres corresponding to the two oscillators coincide as the coupling becomes very strong. 
In terms of $\bar{r} = \displaystyle{\lim_{\varepsilon\to \infty}} 
r_{1*} =\lim_{\varepsilon\to\infty} r_{2*}$,  Eq.~(\ref{eq:MagAngl}) leads to
$\bar{r} (\varrho_1 + \varrho_2 - 2\bar{r}^2) =0$,
the (physically relevant) solution of which is $\bar{r} = \sqrt{(\varrho_1+\varrho_2)/{2}}$, being 
the radius of common limiting sphere (see Fig.~\ref{fig:13}). In fact, from 
Eq.~(\ref{eq:ratio}) one can find the explicit functional dependence of $\rho$ on $\varepsilon$. 
We conclude that there is a transition from phase synchronization to complete synchronization with 
increasing coupling strength between the two oscillators. 

\begin{center}
\begin{figure}[htp]
      \centering
     {\includegraphics[width=0.4\textwidth]{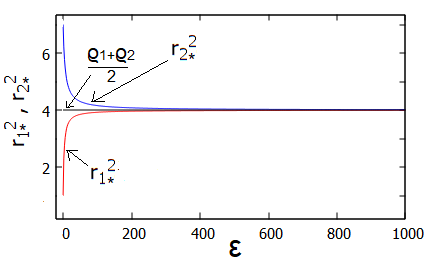}}
     \caption{(Color online) The (squares of the) radii of the limiting spheres for two coupled three-dimensional 
     oscillators of Eq.~(\ref{eq:CpldOsc}), shown as a function of $\varepsilon$, for $\varrho_1=1$ 
     and $\varrho_2=7$. They approach the value $\displaystyle{\lim_{\varepsilon\to \infty}}(r_{1*})^2
     =\lim_{\varepsilon\to\infty}(r_{2*})^2=\tfrac{1}{2}(\varrho_1+\varrho_2)=4$.}
      \label{fig:13}
\end{figure}    
\end{center}

\section{SYMMETRY-BREAKING COUPLING: $D=2$}\label{Ap:5} 
Consider the following general case of two coupled two-dimensional Stuart-Landau oscillators
\begin{align}
\dot{\x}_n &=(\varrho-r_n^2)\x_n +\x_n \cdot \boldsymbol{\mu}^{(n)} + \varepsilon (\x_m-\x_n)\mathsf{R},
\label{eq:SS}
\end{align}
where $\x_n=(x^1_n,x^2_n)$ and $(n.m)=(1,2)$ or (2,1), and $\mathsf{R}$ is a $2\times2$ rotation matrix. Clearly, the system has rotational symmetry since the form of equations remain invariant under rotation. In case the oscillators are identical and $\mathsf{R}$ is a rotation by $\frac{\pi}{2}$, Eq.~(\ref{eq:SS}), in components, leads to
\begin{align}
\dot{x}^1_n &= (\varrho-(r_n)^2)x^1_n + \mu_{12} x^2_n + \varepsilon(x^2_m-x^2_n), \nonumber\\
\dot{x}^2_n &= (\varrho-(r_n)^2)x^2_n - \mu_{12} x^1_n + \eta\varepsilon(-x^1_m+x^1_n),
\label{eq:SB2}
\end{align}
where the trivial factor $\eta=1$ has been introduced for later convenience. For $\varrho>0$, 
the two oscillators synchronize completely, the dynamics being periodic. However, if we make 
$\eta\neq 1$, the equations are no longer symmetric. For instance, for $\eta=-1$, the system 
undergoes oscillation quenching. A more detailed exploration of the effect of $\eta$ on the 
collective dynamics will be an interesting avenue to explore. 

\section{ASYMPTOTICS OF EQS. (\ref{eq:3D2}) AND (\ref{eq:4D2})}\label{Ap:6}
First consider coupling in terms of the eigen-coordinate pair $(y^1_1,y^1_2)$ and $\gamma=1$
in Eqs.~(\ref{eq:3D2}) and (\ref{eq:4D2}). Guided by the dynamics of the uncoupled oscillators, 
we make a reasonable ansatz $\displaystyle{\lim_{t\to \infty}} r_1, r_2 = \sqrt{\varrho}$ for 
both these cases. With this assumption, Eqs.~(\ref{eq:3D2}) and (\ref{eq:4D2}) become linear 
in the asymptotic limit, for which it is easy to see that $\delta_1=y^1_2-y^1_1$ evolve as
$\ddot{\delta}_1 + 2\varepsilon \dot{\delta}_{1} + \omega^2 \delta_{1} = 0$.

For the case of coupled oscillators in $D=3$ [Eq.~(\ref{eq:3D2})] and $D=4$ [Eq.~(\ref{eq:4D2})] 
$\omega = \Omega^{(1)}$ and $\omega_1^{(1)}$, respectively. It follows the above
that $\displaystyle{\lim_{t\to \infty}} \delta_1= 0$, which implies the variables $y^1_2$ and 
$y^1_1$ coincide. Additionally, in case of Eq.~(\ref{eq:4D2}), the equality $y^1_2=y^1_1$ 
necessarily implies $y^2_2=y^2_1$.

Similarly, when coupled in the pair $(y_1^3,y_2^3)$, the difference $\delta_3=y_2^3-y_1^3$ 
corresponding to the three-dimensional oscillators of Eq.~(\ref{eq:3D2}) evolves as
$\dot{\delta}_3 + 2\varepsilon \delta_3=0$,
leading to $\displaystyle{\lim_{t\to \infty}} \delta_3=0$. An identical equation results in the 
case of two coupled four-dimensional oscillators where the parameters have been reduced using 
quaternion algebra, Eq.~(\ref{eq:Q}). When coupled via the pair $(x_1^1,x_2^1)$, the evolution of the 
difference $\delta_x = x_2^1-x_1^1$ (with the condition $\displaystyle{\lim_{t\to \infty}} r_1, r_2 
= \sqrt{\varrho}$), is $\ddot{\delta}_x + 2\varepsilon \dot{\delta}_{x} + \omega^2 \delta_{x} = 0$. 
Hence $\displaystyle{\lim_{t\to \infty}} \delta_x =  0$ leading to partial synchronization.

\end{document}